\documentclass[12pt]{article} 

\usepackage{a4, babel, graphicx, listings, fancyhdr, verbatim}
\usepackage{amsmath}
\usepackage{amsfonts}
\usepackage{amssymb}

\newtheorem{theorem}{Theorem}

\newtheorem{lemma}[theorem]{Lemma}

\newtheorem{proposition}[theorem]{Proposition}

\newenvironment{proof}[1][Proof]{\textbf{#1.} }{\ \rule{0.5em}{0.5em}}

\font\csc=cmcsc10 

\font\bigbfhei=cmbx12 scaled \magstep1
\def\square{{\ \rule{0.5em}{0.5em}}}
\def\hatt{\widehat} 
\def\midd{\,|\,} 
\def\E{{\rm E}} 
\def\d{{\rm d}} 
\def\df{{\rm df}} 
\def\pr{{\rm Pr}} 
\def\N{{\rm N}} 
\def\true{{\rm true}} 
\def\half{\hbox{$1\over2$}}
\def\eps{\varepsilon} 
\def\MCI{\hbox{\csc mci}} 
\def\MSYR{\hbox{\csc msyr}} 
\def\ROI{\hbox{\csc roi}} 
\def\pp{\hbox{\rm piv}}
\def\SSR{\hbox{\csc ssr}}
\def\tr{{\rm t}}
\def\obs{{\rm obs}} 
\def\LR{{\rm LR}} 

\begin{document}
\baselineskip18pt 
\parindent20pt 

\title{Frequentist Analogues of Priors and Posteriors}
\author{{\bigbfhei Tore Schweder} \\ Department of Economics, 
           University of Oslo\ \and 
        {\bigbfhei Nils Lid Hjort} \\ Department of Mathematics, 
           University of Oslo} 
\def\today{February 2002} 
\maketitle
\begin{abstract}
\noindent 
Independent data are efficiently integrated by adding their respective
log-likelihoods. Instead of Bayesian updating of information, we propose to
use the likelihood directly as a vehicle for coherent learning. If past data
are summarised in a likelihood, it might be termed a prior likelihood
component when integrated with new data. In the frequentist tradition,
statistical reporting is often done in the format of confidence intervals. The
confidence distribution, with quantiles specifying all possible confidence
intervals provides a more complete report than a 95\% interval, say. The
concept of confidence distribution is discussed, and a new version of the
Neyman--Pearson lemma is provided. Confidence distributions based on prior data
represent frequentist analogues to Bayesian priors. These confidence
distributions need to be converted to likelihoods before they can be
integrated with the new data likelihood. This can be done if their probability
bases are reported. Confidence distributions resulting from the integrated
analysis, along with their probability bases, represent the frequentist
analogue to the Bayesian posterior distributions.

\smallskip\noindent 
{\csc Key Words:} 
{\sl 
confidence distributions and densities, 
frequentist posteriors and priors, 
integrating information, 
likelihood,
Neyman--Pearson lemma, 
pivots, 
statistical reporting, 
whales } 
\end{abstract}

\section{Introduction}

In econometrics there is a tendency to focus on the
new data in isolation. Other sources of information about the parameters of the
model might enter the analysis in the format of structural assumptions in the
model, possibly as restrictions on certain parameters, or as hypotheses to be
tested, etc. Coherent learning in the Bayesian sense of updating distributional
knowledge of a parameter in the light of new data has no parallel in the
non-Bayesian likelihood- or frequentist tradition. However, the likelihood
function is the pre-eminent tool for integrating diverse data, and provided the
distributional information concerning a parameter has a likelihood
representation, it could simply be integrated with the likelihood of the new
independent data by multiplication.

In the purist likelihood tradition of Edwards (1992) and Royall (1997), 
the likelihood function itself is the primitive. In this tradition, 
the likelihood function of the interest parameter is reported 
and interpreted without recourse to confidence intervals, 
$p$-values or the like. The parallel to Bayesian updating 
is thus straight\-forward. This process of updating a likelihood we term 
\textit{likelihood updating}. When the current knowledge 
of a parameter is represented by a likelihood function, 
this likelihood is updated when combined with the likelihood of new data. 
The ratio between the updated and the previous likelihood 
represents the new information.

The majority of statisticians and econometricians probably
still belong to the frequentist camp. Their favourite format 
of statistical reporting is the confidence interval, 
the $p$-values for certain hypotheses of interest, 
or point estimates accompa\-nied with measures of uncertainty like the
standard error. The likelihood function is, of course, a central concept in
the frequentist tradition, but is mainly used as a tool to obtain efficient
statistics with (asymptotic) frequentist interpretations. Unfortunately, the
statistics reported by frequentist statisticians are seldom sufficient as
input to likelihood updating. To achieve this, the reported statistics have to
be converted to a likelihood function.

We will mainly be concerned with the reporting of frequentist statistics
related to a scalar parameter, and their conversion to a likelihood component.
This does not preclude us from discussing integrative likelihood analysis in
multi-parameter models. As in many Bayesian situations, prior information is
assumed to come in independent packages, one for each one-dimensional
parameter. As the Bayesian, we will also assume that the prior information
comes in the format of a distribution. Instead of regarding the prior
distribution as a probability distribution, we will, however, assume that the
distribution represents confidence intervals by its quantiles. Such
distributions are called confidence distributions, and are studied 
in some detail in Section 2.

Fiducial probability distributions were introduced by Fisher (1930).
In clear-cut cases, fiducial distributions are identical to confidence
distributions. We follow Efron (1998) and others in preferring this latter
term, emphasising the close connection between confidence distributions and
confidence intervals. Following Neyman (1941), the quantiles of the
confidence distribution do, in fact, span confidence intervals. Through 
the bootstrap and other work of Efron (1987, 1993, 1998), 
Fraser (1996, 1998) and others, 
confidence distributions have received renewed interest. 
Schweder and Hjort (2002) present theory for confidence distributions 
and the related reduced likelihoods, and provide more depth and further 
examples to what is presented here.

The well-known duality between hypothesis testing
and confidence intervals emerges as a relationship between the $p$-value 
for a test and the cumulative confidence distribution function. 
In Section 3 a version of the Neyman--Pearson lemma is provided, 
explaining the frequentist optimality of the confidence distribution 
in one-parameter models with monotone likelihood ratio. 
This also leads to optimal constructions of confidence distributions
in parametric families of the exponential kind, via 
conditioning on appropriate statistics. These confidence distributions
become uniformly most powerful in a sense made precise in Section 3.  

We hold that for parameters of primary interest 
the complete confidence distri\-bution should be reported, not only a pair of
quantiles like the endpoints of the 95\% confidence interval. 
We argue in particular that reporting the confidence density
for interest parameters, the derivative of the cumulative 
confidence distribution, is an effective way of summarising results 
from statistical analyses of data, 
regarding both presentation and inference. 
This post-data parameter density curve shares some of the 
features and appealing aspects of the Bayesian posterior, 
but is purely frequentistic, 
and should also be non-controversial conceptually. 

It is desirable to develop methods for obtaining approximate
confidence distri\-butions in situations where exact constructions
either become too intricate or do not exist. In Section 4 
we discuss various approximations, the simplest of which being
based on the traditional delta method for asymptotic normality.
Better versions emerge via corrections of various sorts.
In particular we develop an acceleration and bias corrected bootstrap
percentile interval method for constructing improved 
confidence densities. It has an appealing form and 
is seen to perform well in terms of accuracy. 

For large samples, the asymptotic normality of regular statistics allows
the confidence distribution to be turned into a likelihood function through
its normal scores. This likelihood, called the normal-based likelihood, 
agrees with the so-called implied likelihood of Efron (1993). 
However, if the information content in the data 
is of small to moderate weight, the normal-based
likelihood might be misleading. An example is provided in Section 5 
to show that a given confidence distribution might 
relate to many different likelihood functions,
dependent on the sampling situation behind the confidence distribution. 
For this reason it is advisable to supplement the reported 
confidence distribution with sufficient information 
concerning its probability basis to enable future
readers to recover an acceptable likelihood function related to the 
confidence distribution.

Section 6 develops theory for confidence and likelihoods 
in models with exact or approximate pivots. 
An illustration is given for the problem of 
assessing the stock of bowhead whales outside Alaska in Section 7. 
Finally supplementing remarks and discussion
are found in Section 8. 

There is considerable current interest in building principles for
and evaluating applications of combining different information sources 
in nontrivial situations. Appli\-cation areas range from 
economics to assessments of whale stocks. 
Schweder and Hjort (1997) introduced likelihood synthesis 
to counter problems encountered with earlier attempts at 
Bayesian synthesis, and reviewed part of the literature. 
Recent articles on this theme include Berger, Liseo and Wolpert (1998) 
on eliminating nuisance parameters via integrated likelihoods
as well as Poole and Raftery (1998) on `Bayesian melding';
see also comments in Sections 7 and 8 below. 
Some discussion of bootstrap likelihoods and likelihoods based 
on confidence sets can be found in Davison and Hinkley (1997, Ch.~10). 
Schweder and Hjort (2002) take the material of the present chapter 
further, and discuss the extent to which the methodology 
may be seen as an integrated Fisher--Neyman approach, with 
Fisher (1930) as an important starting point. 

Fisher's fiducial distribution has been controversial, partly because Fisher
insisted that the fiducial distributions are probability distributions on
line with all other probability distributions. Fisher is regarded as the
most important statistician since Laplace and Gauss. Hald (1998) actually
characterises Fisher's work as a paradigmatic revolution on par with that of
Laplace (inverse probability) and Gau\ss{} and Laplace (least squares). 
The controversy around the fiducial distribution has caused some authors 
calling it ``Fisher's biggest blunder'', see Efron (1998). 
This is not the view of Efron, who interprets the fiducial distributions 
as distributions of confidence and not of probability. He says 
(op.~cit.), ``I believe that objective Bayes methods will develop 
for such problems, and that something like fiducial inference
will play an important role in this development.
Maybe Fisher's biggest blunder will become a big hit in the 21st century!''

\section{Confidence distributions}

Before relating confidence distributions to likelihoods, it is worthwhile 
having a closer look at the concept as a format of reporting 
statistical inference.

\subsection{Confidence and statistical inference} 

In a selected and validated econometric model statistical
inference is, to a large extent, carried out as follows. 
From optimality or structural considerations, 
an estimator of the parameter of interest, and possibly 
of the remaining (nuisance) parameters in the model, is determined.
Then, the sampling distribution of the estimator is calculated, possibly by
bootstrapping. Finally, statements of inference, e.g.~confidence intervals,
are extracted from the sampling distribution 
and its dependence on the parameter.

Our context is a parametric model with an interest parameter $\psi$ for which
inference is sought. The interest parameter is assumed to be scalar, and to
belong to a finite or infinite interval on the real line. The space of the
parameter is thus linearly ordered. With inference we shall understand
statements of the type `$\psi>\psi_{0}$', 
`$\psi_{1}\le\psi\le\psi_{2}$', etc., where $\psi_{0}$, $\psi_{1}$ etc.~are 
values usually computed from the data.
To each statement, we would like to associate how much confidence the data
allow us to have in the statement.

As the name indicates, the confidence distribution is related to confidence
inter\-vals, which are interval statements with the confidence fixed 
\textit{ex ante}, and with endpoints calculated from the data. 
A one-sided confidence interval with (degree of) confidence 
$1-\alpha$ has right endpoint the corresponding quantile 
of the confidence distribution. If $C$ is the
cumulative confidence distribution calculated from the data, the left-sided
confidence interval is $(-\infty,C^{-1}(1-\alpha))$. A right-sided confidence
interval $(C^{-1}(\alpha),\infty)$ has confidence $1-\alpha$, 
and a two-sided confidence interval $[C^{-1}(\alpha),C^{-1}(\beta)]$ 
has confidence $\beta-\alpha$. Two-sided confidence
intervals are usually equitailed in the sense that $\alpha=1-\beta$. 

Hypothesis testing and confidence intervals are closely related. Omitting the
instructive proof, this relation is stated in the following lemma.

\begin{lemma}
{\sl 
The confidence of the statement {\rm `$\psi\leq\psi_{0}$'} 
is the cumulative confidence distribution function value 
$C(\psi_{0})$, and is equal to the $p$-value of a test of 
$H_{0}\colon \psi\leq\psi_{0}$ versus 
the alternative $H_1\colon\psi>\psi_{0}$.} 
\end{lemma}

The opposite statement `$\psi>\psi_{0}$' has confidence $1-C(\psi_{0}).$
Usually, the confidence distributions are continuous, and 
`$\psi\geq\psi_{0}$' has the same confidence as `$\psi>\psi_{0}$'.

When $\psi_{0}$ is fixed, the statement `$\psi\neq\psi_{0}$' should, 
preferably, have confidence given by
one minus the $p$-value when testing $H_{0}\colon\psi=\psi_{0}$. 
This can be calculated from the observed confidence distribution, and is 
$1-2\min\{C(\psi_{0}),1-C(\psi_{0})\}$. 

Confidence intervals are invariant w.r.t.~monotone transformations.
This is also the case for confidence distributions. 

\begin{lemma}
{\sl 
Confidence distributions based essentially on the same statistic are in\-variant
with respect to monotone continuous transformations of the parameter: 
If $\rho=r(\psi)$, say, with $r$ increasing, and if $C^{\psi}$ 
is based on $T$ while $C^{\rho}$ is based on $S=s(T)$ 
where $s$ is monotone, then
$$ C^{\rho}(\rho)=C^{\psi}(r^{-1}(\rho)). $$ 
}
\end{lemma}

A sharp distinction should be drawn between the (estimated) sampling
distri\-bution and the confidence distribution. The sampling distribution of the
estimator is the \textit{ex ante} probability distribution of the statistic
under repeated sampling, while the confidence distribution is calculated
\textit{ex post} and distributes the confidence the observed data allow to be
associated with different statements concerning the parameter. Consider the
estimated sampling distribution of the point estimator $\hatt{\psi}$, say
as obtained from the parametric bootstrap. If $\psi^{\ast}$ is a random
estimate of $\psi$ obtained by the same method, the estimated sampling
distribution is the familiar
\[
S(\psi)=\pr\{\psi^{\ast}\leq\psi\mid\hatt{\psi}\}=F_{\hatt{\psi}}(\psi).
\]

The confidence distribution is, in fact, an inversion of a sampling
distribution. It is a distribution for the parameter and not for a
stochastic variable. \textit{Ex ante}, the confidence distribution is a
stochastic element having a probability distribution. 
\textit{Ex post}, however, it is a distribution for the parameter 
calculated from the observed data. Example 1 below illustrates the inversion. 
The inversion is also illustrated by several other examples, 
and is formalised in (2) where the confidence distribution 
is found from a pivot.

%

Exact confidence distributions are 
equivalent to the infamous fiducial distri\-butions 
in the sense of Fisher, at least in cases where Fisher would have
considered the mechanism behind the confidence limits 
to be inferentially correct; 
see the discussion in Efron (1998, Section 8). In view of old
and on-going controversies and confusion surrounding this theme of Fisher,
and the fact that such fiducial distributions sometimes have been 
put forward in ad hoc fashions and with vague interpretation, 
we emphasise that our distributions of confidence are actually 
derived from certain principles in a rigorous framework, 
and with a clear interpretation. 
Our work can perhaps be seen as being in the spirit of Neyman (1941).
We share the view expressed in Lehmann (1993) that 
the division between the Fisherian and the Neyman--Pearson 
tradition is unfortunate. The unity of the two traditions
is illustrated by our version of the Neyman--Pearson lemma 
as it applies to Fisher's fiducial distribution (confidence distribution).
Note also that we in Section 3, in particular, 
work towards establishing confidence distributions 
that are inferentially correct. 

\smallskip
{\bf Example 1.} 
Consider the exponentially distributed variate $T$ with
probability density 
$f(t;\psi)=(1/\psi)\exp(-t/\psi)$. 
From the obvious p-value, the cumulative confidence distribution is 
$C(\psi;t_{\obs})=\pr_\psi\{T>t_\obs\}=\exp(-t_{\obs}/\psi)$. 
\textit{Ex ante}, the potentially observed value $t_{\obs}=T$ 
is stochastic, while \textit{ex post} it is simply an observed value. 
We will often write $C(\psi)$, which is the stochastic quantity 
$C(\psi;T)$ or the realised confidence distribution 
$C(\psi;t_{\obs})$. The interpretation will be clear from the context.
The cumulative confidence distribution function for $\psi$ is 
$C(\psi; t)=\exp(-t/\psi)$. The confidence density is thus 
$c(\psi; t)=(\d/\d\psi)C(\psi; t)=t\psi^{-2}\exp(-t/\psi)$, 
which not only has a completely different interpretation 
from the sampling density of the maximum
likelihood estimator, $T$, but also has a different shape.
\square 

\smallskip
{\bf Example 2.} 
Suppose the ratio $\psi=\sigma_2/\sigma_1$ between
standard deviation para\-meters from two different 
data sets are of interest, where independent estimates 
of the familiar form $\hatt\sigma_j^2=\sigma_j^2 W_j/\nu_j$
are available, where $W_j$ is a $\chi^2_{\nu_j}$. 
The canonical intervals, from inverting the optimal tests
for single-point hypotheses $\psi=\psi_0$, take the form 
$$[\hatt\psi/K^{-1}(1-\alpha)^{1/2},\, 
   \hatt\psi/K^{-1}(\alpha)^{1/2}], $$
where $\hatt\psi=\hatt\sigma_2/\hatt\sigma_1$ and 
$K=K_{\nu_2,\nu_1}$ is the distribution function for 
the $F$ statistic $(W_2/\nu_2)/(W_1/\nu_1)$. Thus 
$C^{-1}(\alpha)=\hatt\psi/K^{-1}(1-\alpha)^{1/2}$.
This corresponds to the confidence distribution function
$C(\psi\midd{\rm data})=1-K(\hatt\psi^2/\psi^2)$, 
with confidence density 
$$c(\psi\midd{\rm data})=k(\hatt\psi^2/\psi^2)2\hatt\psi^2/\psi^3, $$
expressed in terms of the $F$ density $k=k_{\nu_2,\nu_1}$. 
See also Section 3.2 for an optimality result of the 
confidence density used here, and Section 4.3 
for a very good approxi\-mation based on bootstrapping. 
\square 

\smallskip
The calculation of the confidence distribution is easy when a 
pivot statistic for $\psi$ is available. 
The random variable $\pp(X,\psi)$ is a pivot
(Barndorff-Nielsen and Cox, 1994) in a model with nuisance parameter 
$\chi$ and data $X$ if the probability distribution of $\pp(X,\psi)$ 
is the same for all $(\psi,\chi)$, and, secondly, 
the function $\pp(x,\psi)$ is monotone and increasing in $\psi$ for
almost all $x$.

When the confidence distribution is based on a pivot, and $F$ is the
cumulative distribution function of the pivot, the confidence distribution is
\begin{equation}
C(\psi)=F(\pp(X,\psi)). \label{pivot}
\end{equation}
This can also be turned around. If, in fact, $C(\psi;X)$ is a cumulative
confidence distribution based on data $X$, then it is a pivot since at $\psi$
it is uniformly distributed. Thus, a confidence distribution based on a
sufficient statistic exists if and only if there is a pivot based on the
sufficient statistic. And the cumulative confidence distribution function is
simply the probability transform of the pivot.

\subsection{Linear regression}

In the linear normal model, the $n$-dimensional data $Y$ of the response is
assumed $\N(X\beta,\sigma^2 I)$. With $\SSR$ being the residual sum of squares
and with $p={\rm rank}(X)$, $S^{2}=\SSR/(n-p)$ 
is the traditional estimate of the residual variance. 
With $S_{j}^{2}$ being the mean-unbiased estimator of the variance
of the regression coefficient estimator $\hatt{\beta}_{j}$, 
\[
V_j=(\hatt{\beta}_{j}-\beta_{j})/S_j 
\]
is a pivot with a $t$-distribution of $\nu=n-p$ degrees of freedom. 
Letting $t_\nu(\alpha)$ be the quantiles of this $t$-distribution, 
the confidence quantiles for $\beta_{j}$ are
the familiar $\hatt{\beta}_{j}+t_\nu(\alpha)S_{j}$. 
The cumulative confidence distribution function for $\beta_{j}$ 
is seen from this to become 
$$C(\beta_j\midd{\rm data})=1-G_\nu((\hatt\beta_j-\beta_j)/S_j)
  =G_\nu((\beta_j-\hatt\beta_j)/S_j), $$
where $G_\nu$ is the cumulative $t$-distribution 
with $\nu$ degrees of freedom. Note also that the 
confidence density $c(\beta_j\midd{\rm data})$ 
is the $t_\nu$-density centred at $\hatt\beta_j$
and with the appropriate scale. 

Now turn attention to the case where $\sigma$, 
the residual standard deviation, is the parameter of interest. 
Then the pivot $\SSR/\sigma^2=\nu S^2/\sigma^2$ is a $\chi^2_\nu$, 
and the cumulative confidence distribution is found to be 
$$C(\sigma\midd{\rm data})=\pr\{\chi^2_\nu > \SSR/\sigma^{2}\}
  =1-\Gamma_\nu(\nu S^2/\sigma^{2}), $$
where $\Gamma_\nu$ is the cumulative distribution function  
of the chi-square with density $\gamma_\nu$. 
The confidence density becomes 
$$c(\sigma\midd{\rm data})
  =\gamma_\nu\Bigl({\nu S^2\over \sigma^2}\Bigr){2\nu S^2\over \sigma^3} 
  ={S^\nu\over 2^{\nu/2}\Gamma(\half\nu)}
    \sigma^{-(\nu+1)}\exp(-\half\nu S^2/\sigma^2), $$
which again is different from the likelihood. The likelihood,
for the $\SSR$ part of the data, is the density of 
$\SSR=\sigma^2\chi^2_\nu$, which is proportional to
$$L(\sigma)=\sigma^{-\nu}\exp(-\half\nu S^2/\sigma^2). $$
Taking logarithms, the pivot is brought on an additive scale, 
$\log S-\log\sigma$, and in the parameter $\tau=\log\sigma$
the confidence density is proportional to the likelihood. 
The log-likelihood also has a nicer shape in $\tau$ 
than in $\sigma$, where it is less neatly peaked. 

\subsection{Ratio parameters and the Fieller solution}

Ericsson, Jansen, Kerbesian and Nymoen (1998) discuss the empirical basis
for the weights in monetary conditions indices, $\MCI$. 
With $R$ being the interest rate and $e$ the exchange rate, 
$\MCI=\psi (R-R_{0})+(e-e_{0})$.
The relative weight of the interest rate, $\psi$, is estimated as the ratio
of two regression coefficients, 
$\hatt\psi=\hatt\beta_1/\hatt\beta_2$ in a linear regression. 
Both regression parameters are assumed nonnegative, and thus 
$0\le\psi\le\infty$. For Norway, Ericsson et al.~(1998) found 
$[0,\infty]$ as the 95\% confidence interval for $\psi$. 
It is an embarrassing 
conceptual problem to have a 95\% confidence interval covering 
the whole range of the parameter, which certainly should have 
had confidence 100\% and not 95\%. This is known as
the Fieller problem, see e.g.~Koschat (1987) and Dufour (1997). 

Ratios of regression coefficients are often estimated in econometrics. 
The method of Fieller (1940, 1954), discussed below, should therefore 
be of interest. Staiger, Stock and Watson (1997) use the Fieller method to
estimate the non-acceleration rate of unemployment (NAIRU). They found 
in a simulation study that the Fieller method performs much better 
than the delta method. This is not surprising, since this method 
of constructing a confidence region for $\psi$ is related to 
a $t$-test with the usual optimality properties, 
see Neyman's discussion of Fieller (1954).

Assume $(\hatt\beta_1,\hatt\beta_2)^\tr$ to be 
$\N((\beta_1,\beta_2)^\tr,\Sigma)$, with 
$\Sigma =\sigma^2\Sigma_0$ estimated at $\df$ degrees of freedom. 
Confidence regions for the quotient $\psi=\beta_1/\beta_2$ 
are found from inverting the $t$-test of $H_{0}\colon\psi=\psi_0$ 
versus $H_1\colon\psi\not=\psi_0$. The hypothesis is first 
reformulated to $H_0\colon\beta_1-\psi_0\beta_2=0$, 
and tested by $T(\psi_0)$, where  
\begin{equation*}
T(\psi)=\frac{\psi\hatt\beta_2-\hatt\beta_1}
{\hatt\sigma\sigma_0(\psi)}
\quad {\rm and} \quad 
\sigma_0^2(\psi)=(-1,\psi)\Sigma_0(-1,\psi)^{\tr}.
\end{equation*}
This test statistic $T$ is $t$-distributed with $\df$ degrees of freedom
when $\psi=\psi_0$. It is, however, not a pivot since it is not
monotone in $\psi$ for all possible data. Over 
$[-\infty,\infty]$, $T$ typically has two monotone branches. 
With $b=(\hatt\beta_2,-\hatt\beta_1)\Sigma_0$, the point of branching 
is $\psi_m=-b_1/b_2$, which is a minimum if $b_2<0$. 
The inversion of the two-sided test leads to the confidence regions 
of Fieller (1940, 1953). 
For a given level of confidence $1-2\alpha$, the region is determined by 
$\alpha\le G_{\df}(T(\psi))\le 1-\alpha$ where $G_\df$ is
the cdf of the $t_\df$ distribution.
Depending on the confidence level and the estimates, the confidence region
can be either 
$[-\infty,\infty]$, 
$[-\infty,\hatt\psi_U]$, 
$[\hatt\psi_L,\infty]$,
$[\hatt\psi_L,\hatt\psi_U]$, 
or even 
$[\hatt\psi_{L1},\hatt\psi_{L2}] \cup [\hatt\psi_{U1},\hatt{\psi}_{U2}]$ 
with $\hatt\psi_{L2}<\hatt\psi_{U1}$ 
and one or both interval pieces being infinite. 
Note that the coverage probability is exact. 

The Fieller test statistic is $\hatt\beta_1-\psi_0\hatt\beta_2$ 
and orders data differently from the more natural test statistic 
$|\hatt\psi-\psi_0|$. One might therefore suspect that something is lost
by the Fieller solution. However, Koschat (1987) found that no reasonable
method other than the Fieller solution provides confidence intervals with
exact coverage probabilities when $\Sigma_0=I$. 
This was shown for the angle related to the ratio. 
Consequently, if something is to be gained in power 
by another ordering of data sets, it must be at the price of only
obtaining approximate null behaviour.

Since $T(\psi)$ is not a pivot, standard confidence distributions for 
$\psi$ are not available by the Fieller method. 
The method has essentially three problems: 
(i) the confidence region can be infinite; 
(ii) it can be a union of two intervals where at least one is infinite; and 
(iii) it might, at some given degree of confidence
less than 100\%, lead to nonsense interval statements of the type 
$\psi\in[-\infty,\infty]$. 
That the confidence region might be infinite 
is due to the discontinuity of $\psi=\tan(\theta)$ 
at $\theta=\pi/2$. In polar coordinates, the Fieller method 
leads to either an interval for $\theta$ or to the whole half-circle 
$\hatt\theta\pm \pi /2$. If the region for $\theta$ 
is taken to be the fixed interval $(-\pi /2,\pi /2]$, 
the confidence region might consist of two intervals. 
The final, and for some, the most disturbing problem is (iii)
that the confidence region should have 100\% confidence and not 95\% 
when it consists of the complete range of the parameter. 
Following Neyman when discussing Fieller (1954), 
we do however not see this as a disturbing phenomenon. 
Seen as as family of confidence intervals on the circle, the
Fieller method determines a confidence distribution of mass less than 1.
This is indeed possible. In such cases the total mass of the 
confidence distribution determines an upper limit on the confidence 
that can be associated with any informative statement. 
If this mass is less than 95\%, it seems inappropriate to 
present the non-informative 95\% region. One should rather 
reduce the ambition and report what confidence statements the data allow.

\smallskip
{\bf Example 3.} 
For Norway, Ericsson, Jansen, Kerbesian and Nymoen (1998)
found estimates $\hatt\beta_1=0.34891$, $\hatt\beta_2=0.16264$, 
$\sigma_0^2(\psi)=215.998-2\cdot7.992\psi+57.935\psi^2$, 
and $\hatt\sigma=0.01162$. For these data, $T$ is
increasing over the positive axis. 
The transform $C(\psi)=G_{\df}(T(\psi))$ can thus be
taken as the cumulative confidence distribution. 
The cumulative confidence starts at $C(0)=0.025$ and
approaches $C(\infty)=0.962$. Thus, no informative statements with
confidence more than 0.937 are possible for these data.
The confidence distribution represented as the family of equal-tailed
confidence intervals is shown in Figure~\ref{ericssonfiellerintervals}. 

\renewcommand{\figurename}{{\csc Figure}}
\def\hei{-0.0cm} 
\begin{figure} 
\begin{center}
\hspace*{\hei}
\includegraphics[scale=0.66,angle=90]{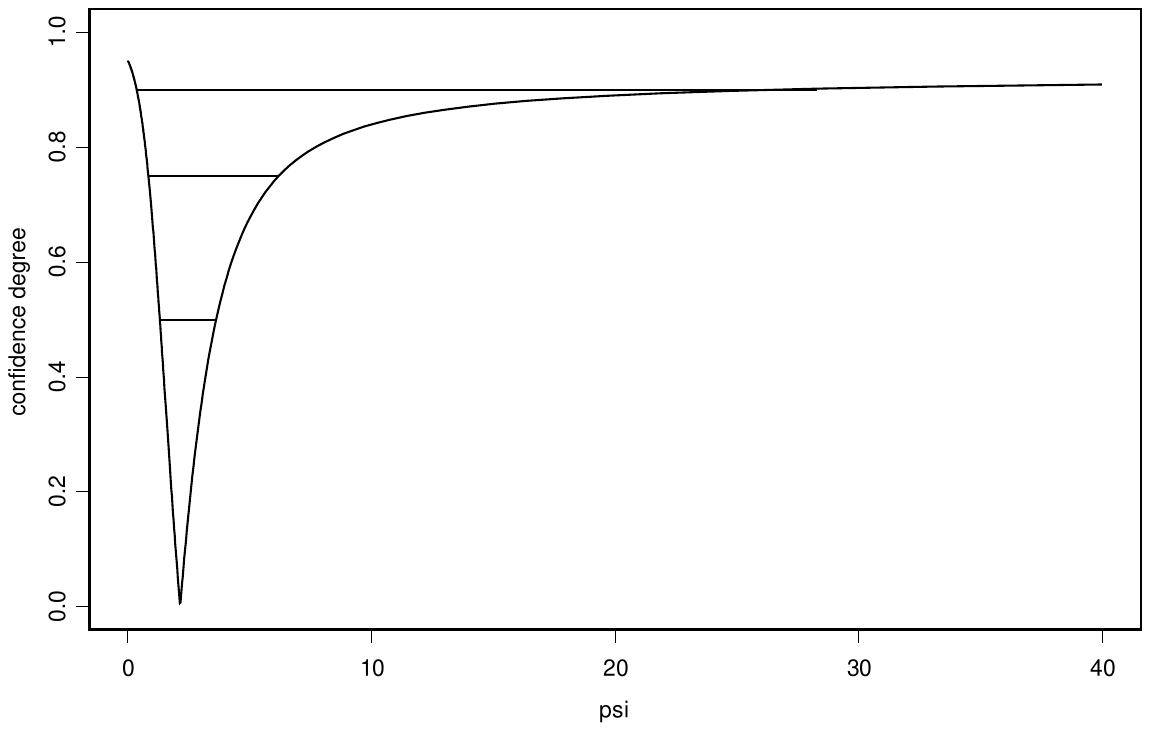}
\caption{\sl Confidence distribution for the
ratio $\protect\psi$ represented as confidence degree
$2\{C(\psi)-\half\}$, based on Norwegian data 
(Nymoen, personal communication).
Equal-tailed confidence intervals at degree 0.50, 0.75 and 0.90
are shown as horizontal lines.}
\label{ericssonfiellerintervals}
\end{center}
\end{figure}



\section{Confidence level and confidence power}

Let $C(\psi)$ be the cumulative confidence distribution. The intended
interpretation of $C$ is that its quantiles are endpoints of confidence
intervals. 
For these intervals to have correct coverage probabilities, 
the cumulative confidence at the true value of the parameter 
must have a uniformly probability distribution. 
This is an \textit{ex ante} statement. 
Before the data have been gathered, the confidence distribution 
is a stochastic element and $C(\psi_{\true})$ is a random variable. If
\begin{equation}
\pr_{\psi}\{C(\psi)\leq c\}=c 
   \quad {\rm for\ }0\leq c\leq1, \label{uniform}
\end{equation}
assuming the probability distribution to be continuous, the coverage
probability of $(-\infty,C^{-1}(\alpha)]$ is
\[
\pr_{\psi}\{\psi\leq C^{-1}(\alpha)\}=\alpha,
\]
which conventionally is called the confidence level of the interval. The
confidence distribution is exact if (\ref{uniform}) holds exactly, and thus
the coverage probability of a confidence interval obtained from $C$ equals the
nominal confidence level.

Confidence distributions provide point estimates, the most natural being the
confidence median, $\hatt{\psi}=C^{-1}(\half)$. When the confidence
distribution is exact, this point estimator is median-unbiased.
This property is kept under monotone transfor\-mations of the parameter. 

The choice of statistic on which to base the confidence distribution is
unambiguous only in simple cases. Barndorff-Nielsen and Cox (1994) are in
agreement with Fisher when emphasising the structure of the model and the data
as a basis for choosing the statistic. They are primarily interested in the
logic of statistical inference. In the tradition of Neyman and Wald, emphasis
have been on inductive behaviour, and the goal has been to find methods with
optimal frequentist properties. In nice models like exponential families 
it turns out that methods favoured on structural and logical grounds also are
favoured on grounds of optimality. This agreement between the Fisherian and
Neyman--Wald schools is encouraging and helps to reduce the distinction between
the two schools. This core of statistical theory needs to be reformulated in
terms of confidence distributions.

\subsection{Optimal confidence power}

The Neyman--Pearson lemma is regarded as a cornerstone of frequentist
theory (cf.~Lehmann, 1959). It says that tests based on sufficient 
statistics with monotone likelihood ratio are uniformly most powerful. 
The power function of a test is an \textit{ex ante} property. 
In the frequentist tradition, methods are judged by their 
\textit{ex ante} properties such as level and power for tests 
and inclusion-exclusion probabilities for confidence intervals. 
The null-property of a confidence distribution is simply (4), 
that $C(\psi)$ has a uniform probability distribution under $\psi$. 
The power properties of $C$ are captured by the probability 
distributions of chosen functionals of $C$. These functionals 
will typically measure the spread of the confidence distribution.

The tighter the confidence intervals are, the better, provided they have the
claimed confidence. \textit{Ex post}, it is thus desirable to have as little
spread in the confidence distribution as possible. Standard deviation, 
inter-quantile difference or other measures of spread could be used 
to rank methods with respect to their discriminatory power. 
The properties of a method must be assessed \textit{ex ante}, 
and it is thus the probability distribution of a
chosen measure of spread that would be relevant. The assessment of the
information content in a given body of data is, however, another matter, and
must clearly be discussed \textit{ex post}. 

When testing $H_{0}\colon\psi=\psi_{0}$ versus $H_{1}\colon\psi>\psi_{0}$, 
one rejects at level $\alpha$ if
$C(\psi_{0})<\alpha.$ The power of the test is 
$\pr\{C(\psi_{0})<\alpha\}$
evaluated at a point $\psi_{1}>\psi_{0}$. Cast in terms of $p$-values, 
the power distribution is the distribution at $\psi_{1}$ of the 
$p$-value $C(\psi_{0})$.
The basis for test-optimality is monotonicity in the likelihood ratio based on
a sufficient statistic, $S$, 
\begin{equation}
\LR(\psi_{1},\psi_{2};S)=L(\psi_{2};S)/L(\psi_{1};S)
\text{ is increasing in }S\text{ \ for }\psi_{2}>\psi_{1}. \label{lik.ratio}
\end{equation}
\def\stoch{{\rm st}}
When $\Pr\{X\le x\}\le \Pr\{Y\le x\}$ for all values $x$, 
$Y$ is stochastically smaller (no larger) than $X$, 
written $Y\le_{\stoch}X$.
From Schweder (1988) we have the following. 

\begin{lemma}
[Neyman--Pearson for $p$-values]{\sl Let $S$ be a one-dimensional 
sufficient statistic with increasing likelihood ratio whenever 
$\psi_{1}<\psi_{2}$. Let the cumulative confidence distribution 
based on $S$ be $C^{S}$ and that based on
another statistic $T$ be $C^{T}$. In this situation,
the cumulative confidence distributions are stochastically ordered: 
$$C^{S}(\psi_{0})\ge_{\stoch(\psi)}C^{T}(\psi_{0}) 
\hbox{{\sl\ at\ }}\psi>\psi_{0}
\quad \hbox{{\sl\ and\ }} \quad 
  C^{S}(\psi_{0})\le_{\stoch(\psi)}C^{T}(\psi_{0}) 
\hbox{{\sl\ at\ }}\psi<\psi_{0}. $$
} 
\end{lemma}

Now, every natural measure of spread in $C$ around the true value of the
parameter, $\psi_{0}$, can be expressed as a functional 
$\gamma(C)=\int_{-\infty}^{\infty}\Gamma(\psi-\psi_{0})\,C(\d\psi)$, 
where $\Gamma(0)=0$, $\Gamma$ is non-increasing to the left of zero, 
and non-decreasing to the right. 
Here $\Gamma(t)=\int_{0}^{t}\gamma(\d u)$ 
is the integral of a signed measure $\gamma$. 

Agree to say that a confidence distribution $C^{S}$ 
is uniformly more powerful in expectation than $C^{T}$ if
\[
\E_{\psi_{0}}\gamma(C^{S})\leq \E_{\psi_{0}}\gamma(C^{T})
\]
holds for all spread-functionals $\gamma$ and at all parameter values
$\psi_{0}$. With this definition, 
the Neyman--Pearson lemma yields the following. 

\begin{proposition}
[Neyman--Pearson for power in the mean]{\sl If $S$ is a sufficient 
one-dimensional statistic and the likelihood ratio {\rm ($\ref{lik.ratio}$)} 
is increasing in $S$ whenever $\psi_{1}<\psi_{2}$, 
then the confidence distribution based on $S$ is
uniformly most powerful in expectation.} 
\end{proposition}

\begin{proof}
By partial integration,
\begin{equation}
\gamma(C)=\int_{-\infty}^{0}C(\psi+\psi_{0})\,(-\gamma)(\d\psi)
+\int_{0}^{\infty}(1-C(\psi+\psi_{0}))\,\gamma(\d\psi). \label{partial}
\end{equation}
By the Neyman--Pearson lemma, 
$\E C^{S}(\psi+\psi_{0})\leq \E C^{T}(\psi+\psi_{0})$ for $\psi<0$ 
while $\E(1-C^{S}(\psi+\psi_{0}))\leq \E(1-C^{T}(\psi+\psi_{0}))$ for $\psi>0$.
Consequently, since both $(-\gamma)(\d\psi)$ and $\gamma(\d\psi)\geq0$,
\[
\E_{\psi_{0}}\gamma(C^{S})\leq \E_{\psi_{0}}\gamma(C^{T}).
\]
This relation holds for all such spread measures that have finite integral,
and for all reference values $\psi_{0}$. Hence $C^{S}$ is 
uniformly more powerful than any other confidence distribution.
\end{proof}

\smallskip
The Neyman--Pearson argument for confidence distributions can be strengthened.
Say that a confidence distribution $C^{S}$ is uniformly most powerful if, 
\textit{ex ante}, $\gamma(C^{S})$ is stochastically less than 
or equal to $\gamma(C^{T})$ for all other statistics, $T$, 
for all spread-functionals $\gamma$, and with respect to 
the probability distribution at all values of the true parameter $\psi_{0}$. 
The following is proved in Schweder and Hjort (2002). 

\begin{proposition}
[Neyman--Pearson for confidence distributions]{\sl If $S$ is a 
sufficient one-di\-mensional statistic and the likelihood ratio 
$(\ref{lik.ratio})$ is increasing in $S$ whenever $\psi_{1}<\psi_{2}$, 
then the confidence distribution based on $S$ is uniformly most powerful.} 
\end{proposition}

\subsection{Uniformly most powerful confidence for exponential families} 

Conditional tests often have good power properties 
in situations with nuisance parameters. In the exponential class
of models it turns out that valid confidence distributions 
must be based on the conditional distribution of the 
statistic which is sufficient for the interest parameter,
given the remaining statistics informative for the nuisance parameters. 
That conditional tests are most powerful among power-unbiased 
tests is well known, see e.g.~Lehmann (1959). There are
also other broad lines of arguments leading to constructions of 
conditional tests, see e.g.~Barndorff-Nielsen and Cox (1994). 
Presently we indicate how and why also the 
most powerful confidence distributions are of 
such conditional nature. 

\begin{proposition}{\sl 
Let $\psi$ be the scalar parameter and $\chi$ the nuisance parameter
vector in an exponential model, with density of the form
$$p(x)=\exp\{\psi S(x)+\chi_1A_1(x)+\cdots+\chi_pA_p(x)
       -k(\psi,\chi_1,\ldots,\chi_p)\} $$
with respect to a fixed measure, for data vector $x$ in a sample space 
region not dependent upon the parameters. 
Assume $(\psi,\chi)$ is contained in an open $(p+1)$-dimensional 
parameter set. Then, for $\psi$ and hence for all monotone transforms 
of $\psi$, there exist exactly valid confidence distributions, 
and the uniformly most powerful of these takes the conditional form
$$C_{S\midd A}(\psi)=\pr_{\psi,\chi}\{S>S_\obs\midd A=A_\obs\}. $$
Here $S_\obs$ and $A_\obs$ denote the observed values of $S$ and $A$. 
}
\end{proposition} 

Strictly speaking the above formula holds in the case 
of continuous distributions; a minor discontinuity correction 
amendment is called for in case of a discrete distribution.

The proof of this proposition, along with extensions and applications, 
are found in Schweder and Hjort (2002). 
One key ingredient here is that $A$ is a sufficient 
and complete statistic for $\chi$ when $\psi=\psi_0$ is fixed. 
Note that the distribution of $S$ given $A=A_\obs$ 
depends on $\psi$ but not on $\chi_1,\ldots,\chi_p$. 

\smallskip
{\bf Example 4.} 
Consider pairs $(X_j,Y_j)$ of independent Poisson variables,
where $X_j$ and $Y_j$ have parameters $\lambda_j$ and $\lambda_j\psi$,
for $j=1,\ldots,m$. The likelihood is proportional to 
$$\exp\Bigl\{\sum_{j=1}^m y_j\log\psi 
  +\sum_{j=1}^m(x_j+y_j)\log\lambda_j\Bigr\}. $$
Write $S=\sum_{j=1}^m Y_j$ and $A_j=X_j+Y_j$. 
Then $A_1,\ldots,A_m$ become sufficient and complete for 
the nuisance parameters when $\psi$ is fixed. 
Also, $Y_j\midd A_j$ is a binomial $(A_j,\psi/(1+\psi))$. 
It follows from the proposition above 
that the uniformly most powerful confidence distribution,
used here with a half-correction for discreteness, takes the simple form 
\begin{align*}
C_{S\midd A}(\psi)&=\pr_{\psi}\{S>S_\obs\midd A_{1,\obs},\ldots,A_{m,\obs}\}
  +\half\pr_{\psi}\{S=S_\obs\midd A_{1,\obs},\ldots,A_{m,\obs}\} \\
&=1-{\rm Bin}\Bigl(S_\obs ; \sum_{j=1}^m A_{j,\obs},
            {\psi\over 1+\psi}\Bigr)
  +\half{\rm bin}\Bigl(S_\obs ; \sum_{j=1}^m A_{j,\obs},
            {\psi\over 1+\psi}\Bigr),  
\end{align*} 
where ${\rm Bin}(\cdot;n,p)$ and ${\rm bin}(\cdot;n,p)$ are 
the cumulative and pointwise distribution functions for the binomial. 
\square 

\section{Approximate confidence distributions}

Uniformly most powerful exact inference is only possible in 
a restricted class of models. 
In more complex models, the statistic upon which to base 
the confidence distribution might be chosen on various grounds: 
the structure of the likelihood function, perceived robustness,
asymptotic properties, computational feasibility, perspective and tradition of
the study. In the given model, with finite data, it might be difficult to
obtain an exact confidence distribution based on the chosen statistic. There
are, however, various techniques available to obtain approximate confidence.

Bootstrapping, simulation and asymptotics are useful tools in calculating
approxi\-mate confidence distributions and in characterising their power
properties. When an estimator, often the maximum likelihood estimator of the
interest parameter, is used as the statistic on which the confidence
distribution is based, bootstrapping provides an estimate of the sampling
distribution of the statistic. This empirical sampling distribution can be
turned into an approximate confidence distribution in several ways. The
simplest and most widely used method of obtaining approximate confidence
intervals is the delta method. This will lead to first order accuracy 
properties in smooth models. A more refined method to obtain 
confidence distributions is via acceleration and bias corrections 
on bootstrap distributions, as developed below. 
This method, along with several other venues for refinement, 
will usually provide second order accuracy properties.

\subsection{The delta method} 

In a sample of size $n$, let the estimator $\hatt{\theta}_n$ 
have an approximate multinormal distribution centred at
$\theta$ and with covariance matrix of the form $S_{n}/n$, so that 
$\sqrt{n}S_{n}^{-1/2}(\hatt{\theta}_{n}-\theta)\rightarrow_d \N(0,I)$. 
By the delta method, the confidence distribution for a parameter 
$\psi=p(\theta)$ is based on linearising $p$ at $\hatt{\theta}$, 
and yields 
\begin{equation}
C_{\rm delta}(\psi)=\Phi((\psi-\hatt{\psi})/\hatt{\sigma}_{n}) 
\label{deltamethod} 
\end{equation}
in terms of the cumulative standard normal. 
The variance estimate is 
$\hatt{\sigma}_{n}^{2}=\hatt g^{\tr}S_{n}\hatt g/n$ 
where $\hatt g$ is the gradient of $p$ evaluated at $\hatt\theta$. 
Again, this estimate of the confidence distribution 
is to be displayed post data with $\hatt\psi$ 
equal to its observed value $\hatt\psi_{\obs}$. 

This confidence distribution is known to be first order
unbiased under weak conditions. That $C_{\rm delta}(\psi)$ is first order
unbiased means that the coverage probabilities converge at the rate
$n^{-1/2}$, or that $C_{\rm delta}(\psi_{\true})$ converges 
in distribution to the uniform distribution at the $n^{1/2}$ rate.
Note also that the confidence density as estimated via the delta method,
say $c_{\rm delta}(\psi)$, is simply the normal density 
$\N(\hatt\psi,\hatt\sigma_n^2)$. 

\subsection{The t-bootstrap method} 

For a suitable monotone transformation of $\psi$ and $\hatt\psi$ 
to $\gamma=h(\psi)$ and $\hatt\gamma=h(\hatt\psi)$, suppose 
\begin{equation}
t=(\hatt\gamma-\gamma)/\hatt\tau
  \quad {\rm is\ an\ approximate\ pivot}, \label{tbootassumption}
\end{equation}
where $\hatt\tau$ is proportional to an estimate of the
standard deviation of $\hatt\gamma$. Let $R$ be the distribution
function of $t$, by assumption approximately independent of 
underlying parameters $(\psi,\chi)$. The canonical confidence 
intervals for $\gamma$ then take the form 
$\hatt\gamma-R^{-1}(1-\alpha)\hatt\tau\le 
  \gamma\le \hatt\gamma+R^{-1}(\alpha)\hatt\tau$, 
which backtransform to intervals for $\psi$, with 
$$C^{-1}(\alpha)=h^{-1}\bigl(\hatt\gamma-R^{-1}(1-\alpha)\hatt\tau\bigr). $$
Solving for $\alpha$ leads to the confidence distribution  
$C(\psi)=1-R((h(\hatt\psi)-h(\psi))/\hatt\tau)$, 
with appropriate confidence density $c(\psi)=C'(\psi)$. 
Now $R$ would often be unknown, but the situation is saved 
via bootstrapping. Let $\hatt\gamma^*=h(\hatt\theta^*)$ 
and $\hatt\tau^*$ be the result of parametric bootstrapping 
from the estimated model. Then the $R$ distribution can be
estimated arbitrarily well as $\hatt R$, say, 
obtained via bootstrapped values of 
$t^*=(\hatt\gamma^*-\hatt\gamma)/\hatt\tau^*$. 
The confidence distribution reported is then as above 
but with $\hatt R$ replacing $R$: 
\begin{equation}
C_{\rm tboot}(\psi)
  =1-\hatt R\bigl((h(\hatt\psi)-h(\psi))/\hatt\tau\bigr). \label{tboot} 
\end{equation} 

This t-bootstrap method applies even when $t$ is not a perfect pivot, 
but is especially successful when it is, since $t^*$ then has 
exactly the same distribution $R$ as $t$. Note that the method
automatically takes care of bias and asymmetry in $R$, and 
that it therefore aims at being more precise than the 
delta method above, which corresponds to zero bias and 
a normal $R$. The problem is that an educated guess is required for a
successful pivotal transformation $h$, and that the interval
is not invariant under monotone transformations. 
The following method is not hampered by these shortcomings. 

\subsection{The acceleration and bias corrected bootstrap method} 

Efron (1987) introduced acceleration and bias corrected 
bootstrap percentile intervals, and showed that these have 
several desirable aspects regarding accuracy and parameter 
invariance. Here we will exploit some of these ideas,
but `turn them around' to construct accurate bootstrap-based
approximations to confidence distributions. 

Suppose that on some transformed scale, from $\psi$ and $\hatt\psi$ 
to $\gamma=h(\psi)$ and $\hatt\gamma=h(\hatt\psi)$, one has 
\begin{equation} 
(\hatt\gamma-\gamma)/(1+a\gamma)\sim\N(-b,1) \label{efron1} 
\end{equation} 
to a very good approximation, for suitable constants $a$ 
(for acceleration) and $b$ (for bias). Both population parameters 
$a$ and $b$ tend to be small, as further commented upon below. 
Assumption (\ref{efron1}) can be written
$\hatt\gamma-\gamma=(1+a\gamma)(Z-b)$,
where $Z$ is standard normal. This leads to 
$$1+a\hatt\gamma=(1+a\gamma)(1+a(Z-b)) $$
and a canonically correct interval for $\gamma$, and hence $\psi$,
as explained in a minute. The essentials of the arguments below 
are that (\ref{efron1}) describes a pivotal model
on a transformed scale and that the apparatus already
established for deriving confidence distributions 
from pivots becomes applicable via the transformation lemma 
of Section 2 in conjunction with bootstrapping. 
We include a little more detail, however, to pinpoint 
the roles of various ingredients. 

Start with $z^{(\alpha)}\le Z\le z^{(1-\alpha)}$,
the symmetric interval including $Z$ with probability $1-2\alpha$,
writing $z^{(\eps)}$ for $\Phi^{-1}(\eps)$. This leads after some
algebra to 
$$h^{-1}\Bigl({\hatt\gamma-(z^{(1-\alpha)}-b)\over 1+a(z^{(1-\alpha)}-b)}\Bigr)
  \le \psi \le 
  h^{-1}\Bigl({\hatt\gamma-(z^{(\alpha)}-b)\over 
      1+a(z^{(\alpha)}-b)}\Bigr). $$
Writing this interval as $[C^{-1}(\alpha),C^{-1}(1-\alpha)]$ 
and solving $C^{-1}(\alpha)=\psi$ for $\alpha$ gives 
the confidence distribution 
\begin{equation}
C(\psi)=\Phi\Bigl({h(\psi)-h(\hatt\psi)\over 1+ah(\psi)}-b\Bigr). 
\label{efr2} 
\end{equation} 
This constitutes a good approximation to the real 
confidence distribution, say $C_{\rm exact}(\psi)$, 
under assumption (\ref{efron1}). But it requires $h$ to be known,
as well as values of $a$ and $b$. 
 
To come around this, look at bootstrapped versions 
$\hatt\gamma^*=h(\hatt\psi^*)$ from the estimated parametric model. 
If assumption (\ref{efron1}) holds uniformly in a neighbourhood 
of the true parameters, then also 
$$(\hatt\gamma^*-\hatt\gamma)/(1+a\hatt\gamma)\midd{\rm data}
  \sim \N(-b,1) $$
with good precision. Hence the bootstrap distribution may be expressed as 
$$\hatt G(t)=\pr_*\{\hatt\psi^*\le t\}
  =\pr_*\{\hatt\gamma^*\le h(t)\}
  =\Phi\Bigl({h(t)-\hatt\gamma\over 1+a\hatt\gamma}+b\Bigr). $$
It follows, again after some algebra, 
that the lower endpoint in the interval for $\psi$ above satisfies  
$$\hatt G(\psi_{\rm low})
  =\Phi\Bigl(b-{z^{(1-\alpha)}-b\over 1+a(z^{(1-\alpha)}-b)}\Bigr), $$
and similarly for $\hatt G(\psi_{\rm up})$. 
This gives firstly the so-called ${\rm BC}_a$ intervals of Efron (1987), 
say $[\hatt C^{-1}(\alpha),\hatt C^{-1}(1-\alpha)]$, 
applying the $\hatt G^{-1}$ transformation here. Secondly 
it gives us an acceleration and bias corrected approximation 
to the confidence distribution found in (\ref{efr2}), 
through solving $\hatt C^{-1}(\alpha)=\psi$ for $\alpha$. 
The result is the abc formula 
\begin{equation}
\hatt C_{\rm abc}(\psi)=\Phi\Bigl(
  {\Phi^{-1}(\hatt G(\psi))-b\over 1+a(\Phi^{-1}(\hatt G(\psi))-b)}-b\Bigr). 
  \label{abcresult}
\end{equation} 
Note that an approximation $c_{\rm abc}(\psi)$ to the 
confidence density emerges too, by evaluating the derivative 
of $\hatt C_{\rm abc}$. This may sometimes be done analytically,
in cases where $\hatt G(\psi)$ can be found in a closed form,
or may be carried out numerically. 

It remains to specify $a$ and $b$. The bias parameter $b$ is found 
from $\hatt G(\hatt\psi)=\Phi(b)$. The acceleration parameter $a$ 
is found as $a={1\over 6}{\rm skew}$, where there are several
ways in which to calculate or approximate the skewness parameter 
in question, at least in i.i.d.~and regression situations. 
Extensive discussions may be found in Efron (1987), 
Efron and Tibshirani (1993, Chs.~14 and 22) and in 
Davison and Hinkley (1997, Ch.~5). One option is via the jackknife method,
which gives parameter estimates $\hatt\psi_{(i)}$ computed 
by leaving out data point $i$, and use 
$$a=(6\sqrt{n})^{-1}{\rm skew}\{\hatt\psi_{(\cdot)}-\hatt\psi_{(1)},
    \ldots,\hatt\psi_{(\cdot)}-\hatt\psi_{(n)}\}. $$
Here $\hatt\theta_{(\cdot)}$ is the mean of the $n$ jackknife estimates. 
Another option for parametric families is to 
compute the skewness of the logarithmic derivative of the 
likelihood, at the parameter point estimate, 
inside the least favourable parametric subfamily;
see again Efron (1987) for more details. 

Note that when $a$ and $b$ are close to zero, the 
abc confidence distribution becomes identical 
to the bootstrap distribution itself. In typical setups, 
both $a$ and $b$ will in fact go to zero with speed 
of order $1/\sqrt{n}$ in terms of sample size $n$. Thus (\ref{abcresult})
provides a second order non-linear correction of shift and scale 
to the immediate bootstrap distribution. 

\smallskip
{\bf Example 5.} Consider again the parameter $\psi=\sigma_2/\sigma_1$
of Example 2. The exact confidence distribution 
was derived there, and is equal to $C(\psi)=1-K(\hatt\psi^2/\psi^2)$,
with $K=K_{\nu_2,\nu_1}$. We shall see how successful the abc apparatus is 
for approximating the $C(\psi)$ and its confidence density $c(\psi)$. 

In this situation, bootstrapping from the estimated 
parametric model leads to $\hatt\psi^*=\hatt\sigma_2^*/\hatt\sigma_1^*$
of the form $\hatt\psi F^{1/2}$, where $F$ has degrees of freedom
$\nu_2$ and $\nu_1$. Hence the bootstrap distribution is 
$\hatt G(t)=K(t^2/\hatt\psi^2)$, and 
$\hatt G(\hatt\psi)=K(1)=\Phi(b)$ determines $b$. 
The acceleration constant can be computed exactly by
looking at the log-derivative of the density $\hatt\psi$, which 
from $\hatt\psi=\psi F^{1/2}$ is equal to 
$p(r,\psi)=k(r^2/\psi^2)2r/\psi^3$. With a little work 
the log-derivative can be expressed as 
$${1\over \psi}\Bigl\{-\nu_2+(\nu_1+\nu_2)
  {(\nu_2/\nu_1)\hatt\psi^2/\psi^2\over 1
   +(\nu_2/\nu_1)\hatt\psi^2/\psi^2}\Bigr\}
  =_d {\nu_1+\nu_2\over \psi} \Bigl\{{\rm Beta}(\half\nu_2,\half\nu_1)
   -{\nu_2\over \nu_1+\nu_2}\Bigr\}. $$
Calculating the three first moments of the Beta gives 
a formula for its skewness and hence for $a$. 
(Using the jackknife formula above, or relatives 
directly based on simulated bootstrap estimates,
obviates the need for algebraic derivations, 
but would give a good approximation only to the $a$ 
parameter for which we here found the exact value.) 

Trying out the abc machinery shows that $\hatt C_{\rm abc}(\psi)$ 
is amazingly close to $C(\psi)$, even when the degrees 
of freedom numbers are low and imbalanced; the agreement
is even more perfect when $\nu_1$ and $\nu_2$ are 
more balanced or when they become larger.
The same holds for the densities $\hatt c_{\rm abc}(\psi)$ and $c(\psi)$;
see Figure \ref{nilsabc}. 
\square 


\renewcommand{\figurename}{{\csc Figure}}
\def\hei{-0.0cm} 
\begin{figure}
\begin{center}
\hspace*{\hei}
\includegraphics[scale=0.66,angle=0]{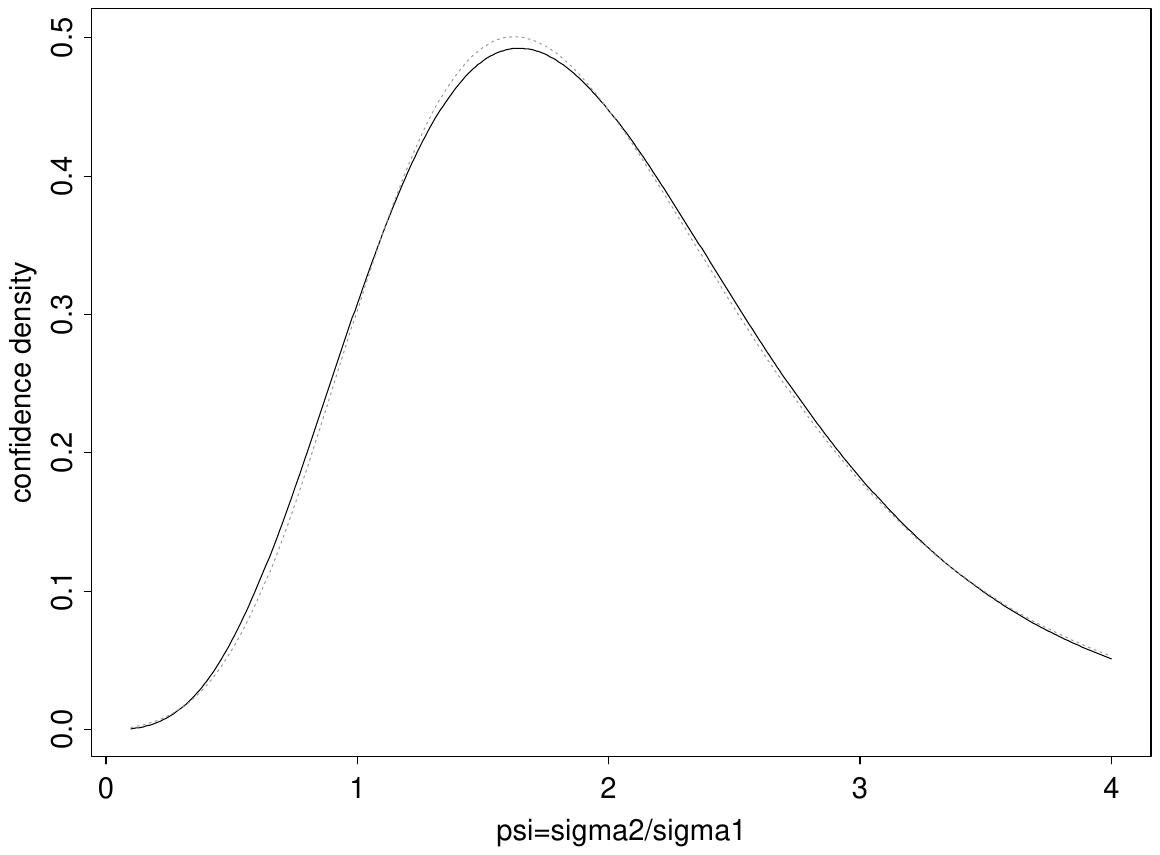}
\caption{\sl True confidence density along with abc-estimated 
version of it, for parameter $\psi=\sigma_2/\sigma_1$
with 4 and 9 degrees of freedom. The parameter estimate 
in this illustration is $\hatt\psi=2.00$.
The agreement is even better when $\nu_1$ and $\nu_2$
are closer or when they are larger.}
\label{nilsabc}
\end{center}
\end{figure}

\subsection{Comments} 

The delta method and the abc method remove bias by transforming 
the quantile function of the otherwise biased normal confidence distribution, 
$\Phi(\psi-\hatt{\psi})$. The delta method simply corrects the scale of
the quantile function, while the abc method applies a shift and a non-linear
scale change to remove bias both due to the non-linearity in $\psi$ as a
function of the basic parameter $\theta$ as well as 
the effect on the asymptotic variance when the basic parameter is changed. 
The t-bootstrap method would have good theoretical properties
in cases where the $\hatt\psi$ estimator is a smooth function
of sample averages, but has a couple of drawbacks compared to 
the abc method. It is for example not invariant under 
monotone transformations. Theorems delineating suitable 
second-order correctness aspects of both the abc and the t-bootstrap 
methods above can be formulated and proved,
with necessary assumptions having to do with the quality
of approximations involved in (\ref{tbootassumption}) and (\ref{efron1}). 
Methods of proof would for example involve Edgeworth 
or Cornish--Fisher expansion arguments; see e.g.~Hall (1992). 
Such could also be used to add corrections 
to the delta method (\ref{deltamethod}). 

Some asymptotic methods of debiasing an approximate confidence distribution
involves a transformation of the confidence itself and not its quantile
function. From a strict mathematical point of view there is of course no
difference between acting on the quantiles or the confidence. But methods like
the abc method above are most naturally viewed 
as a transformation of the confidence for each given value of the parameter.

There are still other methods of theoretical and practical interest 
for computing approximate confidence distributions, cf.~the 
broad literature on constructing accurate confidence intervals. 
One approach would be via analytic approximations to the
endpoints of the abc interval, under suitable assumptions;
the arguments would be akin to those found in DiCiccio and Efron (1996)
and Davison and Hinkley (1997, Ch.~5) regarding 
`approximate bootstrap confidence intervals'. 
Another approach would be via modified profile likelihoods,
following work by Barndorff-Nielsen and others;
see Barndorff-Nielsen and Cox (1994, Chs.~6 and 7) 
and Barndorff-Nielsen and Wood (1998). 
Clearly more work and further illustrations are needed 
to better sort out which methods have the best potential
for accuracy and transparency in different situations. 
At any rate the abc method (\ref{abcresult}) appears 
quite generally useful and precise. 

\section{Likelihood related to confidence distributions}

To combine past reported data with new data, and also for other purposes, it
is advantageous to recover a likelihood function or an approximation thereof
from the available statistics summarising the past data. The question we ask
is whether an acceptable likelihood function can be recovered from a published
confidence distribution, and if this is answered in the negative, how much
additional information is needed to obtain a usable likelihood. 
When a confidence distribution relates to one of many
parameters in the underlying model, the full multi-parameter 
likelihood of the data cannot be obtained from the given one-dimensional 
confidence distribution. What might be recovered is a likelihood reduced 
of these other `nuisance' parameters. This reduced likelihood 
will typically be a marginal or a conditional likelihood.
An example will show that a confidence distribution is in itself 
not sufficient to determine the likelihood of the reduced data,
$T$, summarised by $C$. A given confidence distribution could, 
in fact, result from many different probability models, 
each with a specific likelihood.

Frequentist statisticians have discussed at length how to obtain confidence
distributions for one-dimensional interest parameters from the likelihood of
the data in view of its probability basis. Barndorff-Nielsen and Cox (1994)
discuss adjusted likelihoods and other modified likelihoods based on 
saddle-point approximations, like the $r^{\ast}$. 
Efron and Tibshirani (1993) and Davison and Hinkley (1997) 
present methods based on bootstrapping and
quadratic approximations. These methods are very useful, and in our context
they are needed when inference based on all the available data combined in the
integrated likelihood is done. To obtain the integrative likelihood, the
likelihood components representing the (unavailable) data behind the
confidence distributions are needed, however. To recover a likelihood from a
confidence distribution is a problem that has not been addressed in the
literature, as far as we know.

By definition, a likelihood is a probability density regarded as a function of
the parameters, keeping the data at the observed value. A confidence
distribution can not be interpreted as a probability distribution. 
It distributes confidence and not probability. The confidence density is
therefore not usually a candidate for the likelihood function we seek. It is
the probability distribution of the confidence distribution, regarded as the
data, which matters. We will now demonstrate by means of a 
simple example that a given confidence distribution can relate 
to many different likelihoods, according to its different probability bases.

\smallskip
{\bf Example 6.} 
Consider the uniform confidence distribution over $[0.4,0.8]$. 
The cumulative confidence distribution function is
\begin{equation}
C(\psi)=(\psi-0.4)/0.4 \quad {\rm for\ }0.4\leq\psi\leq0.8. \label{Unif}
\end{equation}
This confidence distribution could have come about in many different ways, and
the likelihood associated with the confidence distribution depends on the
underlying probability models.

\smallskip
{\bf Shift model.} 
In this model, the confidence distribution is based on a statistic with the
sampling property
$$T=\psi-0.2+0.4\,U, $$
where $U$ is uniform (0, 1). The observed value is $T_\obs=0.6$, 
which indeed results in (\ref{Unif}). 
The density of $T$ is $f(t;\psi)=2.5$ on $-0.2<t<\psi+0.2$, 
and the log-likelihood becomes zero on $(T_\obs-0.2,T_\obs+0.2)$ 
and $-\infty$ outside this interval.

\smallskip
{\bf Scale model.} 
The confidence distribution is now based on
$$ T=0.4+(\half\psi-0.2)/U. $$
The observed value $T_\obs=0.6$ leads to the uniform confidence distribution 
over $[0.4,0.8]$. In this scale model the density of $T$ is
$$f(t;\psi)=\frac{\half\psi-0.2}{(t-0.4)^{2}}
  \quad {\rm for\ } t>0.2+\half \psi, $$
and the log-likelihood is $\log(\psi-0.4)$ for $0.4<\psi<2T_\obs-0.4$.

\smallskip 
{\bf Transformed normal model.}
Let now the confidence distribution be based on the statistic
\[
T=0.4\Bigl\{1+\Phi\Bigl(Z+\Phi^{-1}\Bigl(\frac{\psi-0.4}{0.4}\Bigr)
   \Bigr)\Bigr\}\quad {\rm for\ } 0.4\le\psi\le0.8,
\]
where $Z\sim \N(0,1)$. Again, with $T_{\obs}=0.6$, the confidence distribution
is the uniform over the interval $[0.4,0.8]$. 
The likelihood is invariant under a transformation of the data, say
\[
V=\Phi^{-1}(T/0.4-1)=Z+\Phi^{-1}((\psi-0.4)/0.4).
\]
Since then $V_{\obs}=0,$ the log-likelihood is
\[
l(\psi)=-\half \{\Phi^{-1}((\psi-0.4)/0.4)\}^2 
              \quad {\rm for\ }0.4\leq\psi\leq0.8. 
\] 

Three possible log-likelihoods consistent with the uniform confidence
distribution are shown in Figure \ref{3uniform}. 
Other log-likelihoods are also possible.



\renewcommand{\figurename}{{\csc Figure}}
\def\hei{-0.0cm} 
\begin{figure}
\begin{center}
\hspace*{\hei}
\includegraphics[scale=0.66,angle=0]{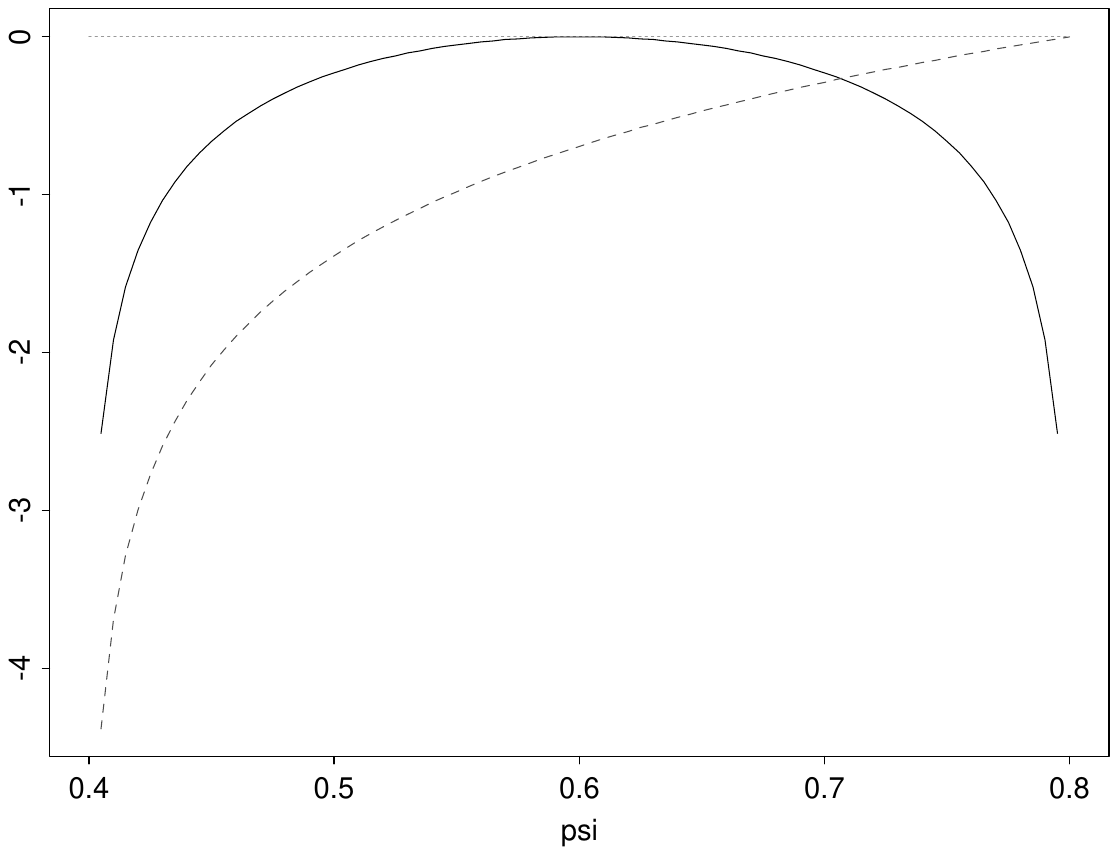}
\caption{\sl Three log-likelihoods consistent with a uniform confidence 
distribution over $[0.4,0.8]$.
{\sl `Many likelihoods informed me of this before, 
which hung so tottering in the balance 
that I could neither believe nor misdoubt.'} -- \csc Shakespeare.}
\label{3uniform}
\end{center}
\end{figure}

\subsection{Confidence and likelihoods based on pivots}


Assume that the confidence distribution $C(\psi)$ is based on a pivot 
$\pp$ with cumulative distribution function $F$ and density $f$. 
Since $\psi$ is one-dimensional, 
the pivot is typically a function of a one-dimensional
statistic $T$ in the data $X$. The probability density of $T$ is then
\begin{equation} 
f^{T}(t;\psi)=f(\pp(t;\psi))\,\Big|\frac{\d\pp(t;\psi)}{\d t}\Big|. 
\label{starone}
\end{equation} 

Here $L(\psi;t)=f^{T}(t;\psi)$ is indeed a likelihood when regarded as a
function of $\psi$ for an observed value of $t$. If $T$ is a sufficient
statistic, $L$ is typically the full likelihood. In other cases, 
when $T$ is a one-dimensional function of a more complex 
sufficient statistic, $T$ is the result of dimension reduction to 
obtain focussed inference for $\psi$. Then the likelihood $L(\psi;t)$ 
is also the result of reducing the dimensionality of the full 
underlying likelihood, and we call it the reduced 
likelihood (Schweder and Hjort, 2002).

Since $\pp(T;\psi)=F^{-1}(C(\psi))$ we have the following. 

\begin{proposition}
{\sl When the probability basis for the confidence distribution is a pivot
$\pp(T;\psi)$ in a one-dimensional statistic $T$, increasing in $\psi$, 
the reduced likelihood is
\[
L(\psi;T)=f\big(F^{-1}(C(\psi))\big)
    \,\Big|\frac{\d\pp(T;\psi)}{\d T}\Big|.
\] }
\end{proposition}

The confidence density is also related to the distribution 
of the pivot. Since one has $C(\psi)=F(\pp(T;\psi))$,
\begin{equation} 
c(\psi)=f(\pp(T;\psi))\,\Big|\frac{\d\pp(T;\psi)}{\d\psi}\Big|. 
\label{startwo} 
\end{equation} 
Thus, the reduced likelihood is in this simple case related to 
the confidence density by
\[
L(\psi;T)=c(\psi)\frac{\d\pp(T;\psi)}{\d T}\left(\Big|
\frac{\d\pp(T;\psi)}{\d\psi}\Big|\right)^{-1}.
\]

There are important special cases. If the pivot is additive in $T$ 
(at some measurement scale), say 
\begin{equation}
\pp(T;\psi)=T-\mu(\psi) \label{locationpivot}
\end{equation}
for a smooth monotone function $\mu$, the likelihood is 
$L(\psi;T)=f\big(F^{-1}(C(\psi))\big)$. 
When furthermore the pivot distribution is normal, we will say
that the confidence distribution has a normal probability basis.

\begin{proposition}
[Normal-based likelihood]{\sl When the probability basis for the confidence
distribution is an additive and normally distributed pivot,
the log-likelihood related to the confidence distribution is
\[
l(\psi)=-\half\left\{\Phi^{-1}(C(\psi))\right\}^2.
\] } 
\end{proposition}

The normal-based likelihood might often provide a good approximate likelihood.
Note that classical first order asymptotics lead to normal-based likelihoods. 
The conventional method of constructing confidence intervals 
with confidence $1-\alpha$,
\[
\{\psi\colon 2(l(\hatt{\psi})-l(\psi)) < \Phi^{-1}(1-\half \alpha)\}
\]
where $\hatt{\psi}$ is the maximum likelihood estimate, 
is equivalent to assuming the likelihood to be normal-based.
The so-called ABC confidence distributions of Efron (1993), 
concerned partly with exponential families, 
have asymptotic normal probability basis, as have confidence distributions 
obtained from Barndorff-Nielsen's $r^{\ast}$ 
(Barndorff-Nielsen and Wood, 1998). 


In many applications, the confidence distribution is found by simulation. 
One might start with a statistic $T$ which, together with an (approximate)
ancillary statistic $A$, is simulated for a number of values of the interest
parameter $\psi$ and the nuisance parameter $\chi$. The hope is that the
conditional distribution of $T$ given $A$ is independent of the nuisance
parameter. This question can be addressed by applying regression methods to
the simulated data. The regression might have the format
\begin{equation}
T=\mu(\psi)+\tau(\psi)V \label{LocationScale}
\end{equation}
where $V$ is a scaled residual. Then
$\pp(T;\psi)=(T-\mu(\psi))/\tau(\psi)$,
and the likelihood is
\[
L(\psi)=f\big(F^{-1}(C(\psi))\big)\big/\tau(\psi).
\]
The scaling function $\tau$ and the regression function $\mu$ 
might depend on the ancillary statistic.


\smallskip
{\bf Example 7.} 
Let $X$ be Poisson with mean $\psi$. The half-corrected cumulative
confidence distribution function is
\[ C(\psi)=1-\sum_{x=0}^{X}e^{-\psi}\frac{\psi^{x}}{x!}
   +\half e^{-\psi}\frac{\psi^{X}}{X!}e^{-\psi}. \]
Here $Y=2(\sqrt{\psi}-\sqrt{X})$ is approximately $\N(0,1)$
and is accordingly approximately a pivot for moderate to large $\psi$. 
From a simulation experiment, one finds that the distribution of 
$Y$ is slightly skewed, and has a bit longer tails than the normal. 
By a little trial and error, one finds that $\exp(Y/1000)$ is closely
Student distributed with $\df=30$. With $Q_{30}$ being the upper quantile
function of this distribution and $t_{30}$ the density, the log-likelihood is
approximately
$$ l_{s}(\psi)=\log t_{30}(Q_{30}(C(\psi)) - \log t_{30}(0). $$
Examples are easily made to illustrate that the 
$l_s(\psi)$ log-likelihood quite closely approximates the 
real Poisson log-likelihood $l(\psi)=x-\psi+x\log(\psi/x)$. 
\square 

\smallskip 
Usually, the likelihood associated with a confidence distribution is different
from the confidence density. The confidence density depends on the
parametrisation. By reparametrisation, the likelihood can be brought to be
proportional to the confidence density. This parametrisation might have
additional advantages. 

Let $L(\psi)$ be the likelihood and $c(\psi)$ the confidence density for
the chosen parametrisation, both assumed positive over the support of the
confidence distribution. The quotient
$$J(\psi)=L(\psi)/c(\psi) $$
has an increasing integral $\mu(\psi)$, with $(\d/\d\psi)\mu=J$, 
and the confidence density of $\mu=\mu(\psi)$ 
is $L(\mu(\psi))$. There is thus always a parametrisation 
that makes the likelihood proportional to the confidence density.

When the likelihood is based upon a pivot of the form $\mu(\psi)-T$, 
the likelihood in $\mu=\mu(\psi)$ is proportional to 
the confidence density of $\mu$.  

\smallskip
{\bf Example 8.} 
Let $\hatt{\psi}/\psi$ be standard exponentially distributed. 
Taking the logarithm, the pivot is brought on translation
form, and $\mu(\psi)=\log\psi$. 
The likelihood and the confidence density is thus 
$c(\mu)\propto L(\mu)=\exp(\hatt{\mu}-\mu-\exp(\hatt{\mu}-\mu))$. 
Bootstrapping this confidence distribution and likelihood is
achieved by adding the bootstrap residuals $\log V^{\ast}$ to $\hatt{\mu}$
above, where $V^{\ast}$ is standard exponentially distributed. The
log-likelihood has a more normal-like shape in the $\mu$ parametrisation than
in the canonical parameter $\psi$. Also, being a translation family in 
$\mu$, the likelihood and the confidence density are easily interpreted.
\square 

\smallskip
Schweder and Hjort (2002) denote the likelihood obtained 
from a pivot (as in Proposition 7) the \textit{reduced likelihood}. 
It is reduced in more than one sense. Nuisance parameters
are removed and the reduced likelihood is a likelihood in
the interest parameter only. The data are reduced to the
scalar statistic in the pivot, and the model is in a sense
reduced to the model for the pivot alone. -- Marginal 
and conditional likelihoods, when they exist, 
are often useful as reduced likelihoods. There is of course 
a substantial literature on likelihoods and their modifications, 
see e.g.~Barndorff-Nielsen and Cox (1994). 

Fisher(1930) introduced his fiducial distributions, here called
confidence distributions, through pivots and (\ref{startwo}). 
It is remarkable that he en passant also gave (\ref{starone}). 
Fisher clearly understood (\ref{starone}) as a likelihood
counterpart to (\ref{startwo}). 
This insight seems not to have survived well in the literature.

\section{Illustration: A population dynamics model\\
 for bowhead whales}

The management of fisheries and whaling rests on the quality of stock
assessments. Within the International Whaling Commission, the assessment of
the stock of bowhead whales subject to exploitation by Native Americans in
Alaska has been discussed repeatedly. The most recent assessment was done by
Bayesian methodology, see International Whaling Commission (1999), 
Poole and Raftery (1998) and Schweder (2002). 

Raftery and his co-workers have developed the Bayesian synthesis approach, 
and in Poole and Raftery (1998) the most recent version of this approach 
is presented. In addition to using their method on the more complex 
age-specific population dynamics model used by the 
International Whaling Commission, they
also illustrate their method on a simplified version of the model. To
illustrate our approach, and to compare it to the Bayesian synthesis approach
with `Bayesian melding' to harmonise conflicting prior distributions, it is
applied to the simplified population dynamics model with the same data as used
by Poole and Raftery (1998, Section 3.6).

Poole and Raftery (1998) understand their prior distributions as prior
probability distributions to be used in a Bayesian integration. We
re-interpret this prior distributions as prior confidence distributions.
Instead of a Bayesian integration, this confidence distributions will be
integrated with the likelihood of $P_{1993}$ and 
yearly rate of increase $\ROI$, after having been
converted to likelihoods. To do this, their probability basis needs to be
determined. As Bayesians, Poole and Raftery are not concerned with this issue,
and no information is thus available. For simplicity, we chose to assume a
normal probability basis for each of the three prior confidence distributions.
This determines their likelihoods as follows.

At the beginning of year $t$, there are $P_{t}$ bowhead whales. With
$C_{t}$ being the catch in year $t$, the deterministic dynamical model is
$$P_{t+1}=P_{t}-C_{t}+P_t\,1.5\,\MSYR
   \{1-(P_t/P_{1848})^2\}, $$
where $P_{1848}$ is taken as the carrying capacity of the stock, and
$\MSYR$ (maximum sustainable yield rate) is the productivity parameter.
Yankee whaling started in 1848. From that year on, the catch history
$\left\{C_{t}\right\} $ is available, and is assumed to be exactly known. 
There are two free parameters in this model. 
We choose these to be $\MSYR$ and $P_{1848}$. 
Stock sizes for other years are also parameters, but are determined by
$\MSYR$ and $P_{1848}$. Poole and Raftery (1998) list 
the following independent priors: 
\begin{align*}
P_{1848}  & \sim6400+\text{Gamma}(2.81,0.000289),\\
\MSYR  & \sim\text{Gamma}(8.2,372.7),\\
P_{1993}  & \sim \N(7800,1300^{2}),
\end{align*}
where Gamma$(a,b)$ denotes the gamma distribution with mean $a/b$ and
standard deviation $\sqrt{a}/b$.

In addition to prior information, there are two
log-likelihood components based on recent survey data. 
One component concerns $P_{1993}$, and is the Gau\ss ian
$$l_{4}(P_{1993})=-\half\{(P_{1993}-8293)/626\}^2. $$
The source of this information is different from that 
of the prior distribution for $P_{1993}$ above.  
The other component concerns the recent rate of increase in stock size
$\ROI$, which is defined through $P_{1993}=(1+\ROI)^{15}P_{1978}$.
It has likelihood
\[
l_{5}(\ROI)=\frac{9}{2}\log\left\{1+\frac{1}{8}\left(  \frac{\log
(1+\ROI)-0.0302}{0.0069}\right)  ^{2}\right\}  -\log(\ROI+1)
\]
obtained from the t-distribution with 8 degrees of freedom, and an exponential
transformation (Poole and Raftery, 1998).

The combined log-likelihood
\[
l=l_{1}(P_{1848})+l_{2}(\MSYR)+l_{3}(P_{1993})+l_{4}(P_{1993})+l_{5}(\ROI)
\]
is an extremely narrow curved ridge. See the bootstrap sample of maximum
likelihood estimate in Figure \ref{bootstrapBowhead}. 
The maximum likelihood estimate is presented in the table 
together with quantiles of the confidence distributions 
for the various parameters obtained by the abc method, 
having employed $1000$ bootstrap replicates.

The bootstrapping is determined by the probability bases of the likelihood
components as follows: 
\begin{align*}
l_{1}^{\ast}(P_{1848})  & =-\half
  \left\{  \Phi^{-1}(G_{2.81,0.000289}(P_{1848}-6400))+Z^{\ast}\right\}^{2}, \\
l_{2}^{\ast}(\MSYR)  & =-\half
  \left\{  \Phi^{-1}(G_{8.2,372.7}(\MSYR))+Z^{\ast}\right\}^{2}, \\
l_{3}^{\ast}(P_{1993})  & =-\half\{(P_{1993}-7800)/1300+Z^\ast\}^2, \\
l_{4}^{\ast}(P_{1993})  & =-\half\{(P_{1993}-8200)/564+Z^{\ast}\}^2, \\
l_{5}^{\ast}(\ROI)  & =\frac{9}{2}\log\left\{  1+\frac{1}{8}\left(  \frac
{\log(1+\ROI)-0.0302}{0.0069}+T_{8}^{\ast}\right)^{2}\right\} -\log(\ROI+1).
\end{align*}
Here the four $Z^{\ast}$s are independently drawn from the standard normal
distribution and independently from $T_{8}^{\ast}$ which is drawn from the
$t_{8}$-distribution. For each draw of these five `bootstrap residuals', 
the perturbed likelihood is maximised, leaving us with the bootstrap estimate
$\theta^{\ast}=(P_{1848}^{\ast},\MSYR^{\ast})$. 

Each bootstrap replicate $\theta^{\ast}$ induces bootstrap replicates of
$P_{1993}$ and $\ROI$ through the deterministic population dynamics model. The
bootstrap distributions are nearly normal for both the two induced parameters,
$P_{1993}^{\ast}\sim \N(8117,556^{2})$ and 
$\ROI^{\ast}\sim \N(0.0249,0.0059^{2})$. 
By simple bias correction relative to the maximum likelihood estimates, we
obtain the quantiles in the table. 
For the primary parameters, transformation
is necessary to obtain normality for the bootstrap sample. 
This is achieved by a Box--Cox transformation of exponent --1: 
$10000/P_{1848}^{\ast}\sim \N(0.7525,0.0748^{2})$, 
while a square root transform normalises the bootstrap distribution of $\MSYR$:
\begin{equation}
(\MSYR^{\ast})^{1/2}\sim \N(0.1611,0.0206^{2}). \label{MSYR*}%
\end{equation}
Kolmogorov--Smirnov tests yielded $p$-values around 0.5 in both cases. 
A simple bias correction on these scales leaves us with 
the confidence quantiles in the table.

\begin{table}[tbp] \centering
\begin{tabular}
[c]{lllllll}%
Parameter & MLE & Quantile & Prior & Posterior & Bayesian\\\hline
$P_{1848}$ &  &0.025 & 8263 & 10932 & 12057\\
& 13152 &0.5 & 14997 & 13017 & 14346\\
&  &0.975 & 30348 & 16085 & 17980\\
&  &  &  &  & \\
$\MSYR$ &  &0.025 & 0.0096 & 0.0157 & 0.0113\\
& 0.0267 &0.5 & 0.0211 & 0.0275 & 0.0213\\
&  &0.975 & 0.0394 & 0.0425 & 0.0333\\
&  &  &  &  & \\
$P_{1993}$ &  &0.025 &  & 7027 & 7072\\
& 8117 &0.5 &  & 8117 & 8196\\
&  &0.975 &  & 9207 & 9322\\
&  &  &  &  & \\
$\ROI$ &  &0.025 &  & 0.0145 & 0.0105\\
& 0.0255 &0.5 &  & 0.0261 & 0.0204\\
&  &0.975 &  & 0.0377 & 0.0318\\\hline
\end{tabular}
\caption{\sl Maximum likelihood estimates and quantiles 
for prior and posterior of frequentist and Bayesian distributions, 
for the parameters of interest. \label{key}}%
\end{table}%

The probability basis of the input confidence distributions leading to the
log-likelihoods $l_{1},l_{2}$ and $l_{3}$ provides a basis for the bootstrap,
as explained above. For $\ROI$, we assume the likelihood given by Poole and
Raftery to be based on the estimate of $\ROI$, appropriately transformed, being
a $T_{8}$ distance from the transformed parameter, where $T_{8}$ is drawn from
a t-distribution with 8 degrees of freedom.


\def\hei{-0.0cm} 
\begin{figure}
\begin{center}
\hspace*{\hei}
\includegraphics[scale=0.66,angle=0]{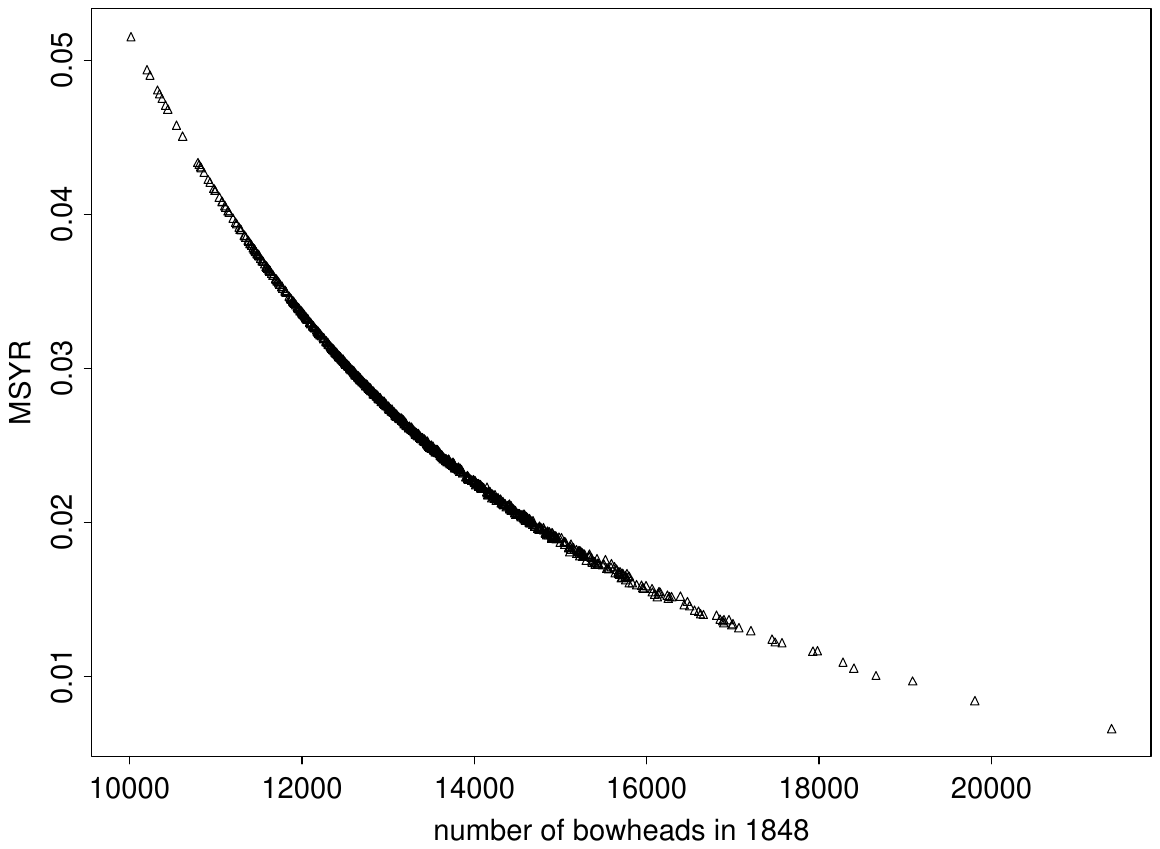}
\caption{\sl One thousand bootstrap estimates of $P_{1848}$ and $\MSYR$.}
\label{bootstrapBowhead}
\end{center}
\end{figure}

The prior and the posterior confidence densities of $\MSYR$ 
are shown in Figure \ref{msyr.plot}. 
The main reason for the posterior being shifted to the right
of the prior confidence distribution is the influence of the data on $\ROI$.
The bootstrap density for $\MSYR$ is also shown. Note that the bias correction
pushed the posterior confidence distribution towards higher values of $\MSYR$.
The bias correction is roughly +6\% in this parameter.

The probability basis for the posterior confidence distribution for $\MSYR$ 
is normal. It is, in fact, based on (\ref{MSYR*}) and an assumption of the
posterior confidence distribution on this scale only being shifted by a
constant amount relative to the bootstrap distribution,
${\MSYR}^{1/2}\sim ({\MSYR^{\ast}})^{1/2}+b$. 
The bias correction $b$ is estimated as 
$$2\bigl[(\hatt\MSYR)^{1/2}
   -{\rm mean}\{(\MSYR^{\ast})^{1/2}\}\bigr]=0.0046, $$ 
where $\hatt\MSYR$ is the maximum likelihood estimate of $\MSYR$. 
With the probability basis being normal, 
the posterior log-likelihood of $\MSYR$ is
$$l_{\rm post}(\MSYR)=-\half(\MSYR^{1/2}-0.166)^2/0.0206^2. $$


\def\hei{-0.0cm} 
\begin{figure}
\begin{center}
\hspace*{\hei}
\includegraphics[scale=0.66,angle=0]{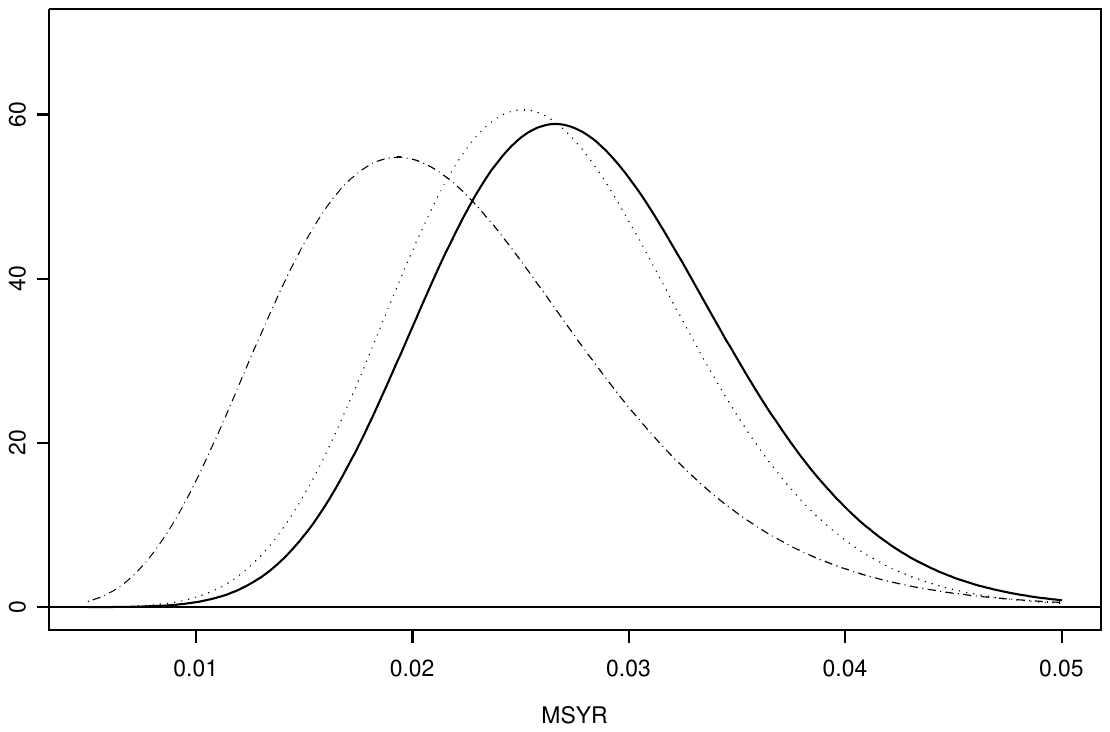}
\caption{\sl Prior confidence density (broken line), bootstrap density (dotted)
and posterior confidence densities for $\MSYR$.}
\label{msyr.plot}
\end{center}
\end{figure}

\section{Discussion}

The confidence distribution is an attractive format for reporting statistical
inference for parameters of primary interest. To allow future good use of the
results it is desirable to allow a likelihood to be constructed from the
confidence distribution. An alternative is to make the original data
available, or to present the full likelihood. 
The work invested in reducing the original data to a confidence 
distribution for the parameter of interest would then be lost, however.
To convert the posterior confidence distribution to a likelihood 
the probability basis for the confidence distribution must be reported. 

Our suggestion is accordingly to extend current frequentist reporting 
practice from only reporting a point estimate, 
a standard error and a (95\%) confidence
interval for the parameters of primary interest. To help future readers, 
one should report the confidence distribution fully, and supplement it with
information on its probability basis, e.g.~the pivot and its distribution.
This latter information might be qualitative.

\subsection{Frequentist updating of information} 

The advantages of representing the information contained in a confidence
distribution in the format of (an approximate) likelihood function are 
many and substantial. 

By adding the log-likelihoods of independent confidence distributions
for the same parameter, an integrated likelihood, and thus a combined
confidence distribution is obtained. The merging of independent confidence
intervals has for example attracted considerable attention, 
and the use of reduced likelihoods presents a solution to the problem. 
One might, for example, wish to merge independent confidence intervals 
for the same parameter to one interval based on all the data. 
When the probability basis and the confidence distribution 
are known for each data set, the related log-likelihoods can be added, 
and an integrative confidence distribution (accompanied with its 
probability basis) is obtained.

A related problem is that of so-called meta-analyses. 
If independent confidence distributions are obtained 
for the same parameter, the information is combined
by adding the accompanying reduced log-likelihoods. A frequent problem in 
meta-analysis is, however, that the interest parameter 
might not have exactly the same value
across the studies. This calls for a model that reflects this variation,
possibly by including a random component. In any event, the availability of
reduced likelihood functions from the various studies facilitates the 
meta-analysis, whether a random component is needed or not.

Studies in fields like ecology, economics, geophysics etc.~often
utilise complex models with many parameters. To the extent results are
available for some of these parameters, it might be desirable to include this
information in the study. If these previous results appear in the format of
confidence distributions accompanied by explicit probability bases, their
related likelihoods are perfectly suited to carry this information into the
combined likelihood of the new and the previous data. If a confidence
distribution is used that is not based on (previous) data, but on subjective
judgement, its related likelihood can still be calculated and combined with
other likelihood components, provided assumptions regarding its probability
basis can be made. This subjective component of the likelihood should then,
perhaps, be regarded as a penalising term rather than a likelihood term.

Finally, being able to obtain the related likelihood from confidence
distributions, and being able to calculate confidence distributions from data
summarised by a likelihood within a statistical model, a methodology parallel
to and competing with Bayesian methodology emerges. This methodology is
frequentist in its foundation. As the Bayesian methodology, it provides a
framework for coherent learning and its inferential product is a distribution:
a confidence distribution instead of a Bayesian posterior probability
distribution. 
To highlight the parallel, the confidence distribution obtained in a
study might be regarded as a frequentist posterior distribution. 
The reduced likelihood, related to an earlier confidence distribution, 
can on the other hand be regarded as a frequentist prior 
when used in a likelihood synthesis.

\subsection{Differences from the Bayesian paradigm} 

It is pertinent to compare the Fisher--Neyman approach,
as further developed in this chapter and in Schweder and Hjort (2002), 
with the Bayesian approach to coherent learning. 
Most importantly, the two approaches have the same aim:
to update distributional knowledge in the view of new data within the frame of
a statistical model. The updated distribution could then be subject to further
updating at a later stage, etc. In this sense, our approach could be termed
`frequentist Bayesian' (a term both frequentists and Bayesians probably
would dislike). There are, however, substantial differences between the two
approaches. Compared to the Bayesian approach, we would like to emphasise 
the following.

Distributions for parameters are understood as confidence distributions
and not probability distributions. The concept of probability is reserved for
(hypothetically) repeated sampling, and is interpreted frequentistically. 
To update a confidence distribution it must be related to 
its probability basis, as the likelihood related 
to the confidence distribution. To update a distribution 
the frequentist needs more information than the Bayesian, 
namely its probability basis.
On the other hand, the distinction between probability
and confidence is basic in the frequentist tradition.

The frequentist can start at scratch, without any (unfounded) subjective
probability distribution. In complex models, there might be distributional
information available for some of the parameters, but not for all. The
Bayesian is then stuck, or she has to construct priors. The frequentist will,
however, not have principle problems in such situations. The concept of
non-informativity is, in fact, simple for likelihoods. The non-informative
likelihoods are simply flat. Non-informative Bayesian priors are, on the other
hand, a thorny matter. In general, the frequentist approach is
less dependent on subjective input to the analysis than the Bayesian approach. 
But if subjective input is needed, it can readily be incorporated (as a
penalising term in the likelihood).

In the bowhead example, there were three priors but only two free
parameters. Without modifications of the Bayesian synthesis approach like the
melding of Poole and Raftery (1998), the Bayesian gets into trouble. Due to
the Borel paradox (Schweder and Hjort, 1997), the Bayesian
synthesis will, in fact, be completely determined by the particular
parametrisation. With more prior distributions than there are free parameters,
Poole and Raftery (1998) propose to meld the priors to a joint prior
distribution of the same dimensionality as the free parameter. This melding is
essentially a (geometric) averaging operation. If, however, there are
independent prior distributional information on a parameter, it seems wasteful
to average the priors. If, say, all the prior distributions happen to be
identical, their Bayesian melding will give the same distribution. The Bayesian
will thus not gain anything from $k$ independent pieces of information, while
the frequentist will end up with a less dispersed distribution; 
the standard deviation will, in fact, be the familiar $\sigma/\sqrt{k}$.

Non-linearity, non-normality and nuisance parameters can produce bias
in results, even when the model is correct. 
The maximum likelihood estimator of variance in a standard regression model
is biased, for example. From a frequentist point of view, 
Bayesian posterior distributions can also be severely biased. 
This has been emphasised repeatedly in the frequentist literature. 
Such bias should, as far as possible, 
be corrected in the reported results. The confidence distribution
aims at being unbiased: when it is exact, the related confidence intervals
have exactly the nominal coverage probabilities. Bias correction has
traditionally not been a concern in the Bayesian tradition. There has,
however, been some recent interest in the matter. To obtain frequentist
unbiasedness, the Bayesian will have to choose her prior with unbiasedness in
mind. Is she then a Bayesian? Her prior distribution will then not represent
prior knowledge of the parameter in case, but an understanding of the model.
Our `frequentist Bayesianism' solves this problem in principle. It takes as
input (unbiased) prior confidence distributions and delivers (unbiased)
posterior confidence distributions.

Efron (1998) regards Bayesians as the optimists amongst statisticians, 
while the frequentists are the pessimists. The Bayesian will
always (in principle) obtain a proper posterior distribution when 
his prior is proper. (When it is improper, as often is the case, 
there might be unforeseen consequences, see e.g.~van Dijk (this volume).) 
The frequentist, however, might end up with an improper 
confidence distribution, as was the case for
Ericsson et al.~(1998), see Section 2.3. The frequentist might, 
in fact, find that the data only allow informative statements 
(finite confidence intervals) up to a certain level of confidence. 
`Life is easier' for the Bayesian also in other respects. 
She has a clear concept of joint posterior distributions, 
and she obtains her one-dimensional posteriors by simple marginalisation. 
Multivariate confidence distributions are difficult. Only
in special cases are they available, and then marginalisation leads to
acceptable one-dimensional confidence distributions when further assumptions
are satisfied. The difficulties stem from the insistence on the
null-property of confidence distributions -- that any confidence interval
shall have the nominal coverage probability. 
Unbiasedness in this sense has its price, usually disregarded by Bayesians.


\end{document}